\documentclass[prd,aps,twocolumn,superscriptaddress,nofootinbib,preprintnumbers,floatfix]{revtex4-1}
\pdfoutput=1

\usepackage{amsmath}
\usepackage{amssymb}
\usepackage{graphicx}
\usepackage[small]{subfigure}
\usepackage[hyperfootnotes=false]{hyperref}
\usepackage{xspace}
\usepackage{xcolor}

\newcommand{\eq}[1]{Eq.~\eqref{eq:#1}}

\newcommand{\fig}[1]{Fig.~\ref{fig:#1}}
\newcommand{\figs}[2]{Figs.~\ref{fig:#1} and \ref{fig:#2}}
\renewcommand{\sec}[1]{Sec.~\ref{sec:#1}}

\newcommand{\ord}[1]{\mathcal{O}(#1)}

\newcommand{\df}{\mathrm{d}}
\newcommand{\img}{\mathrm{i}}

\newcommand{\sdt}{\!\cdot\!}

\newcommand{\al}{\alpha}
\newcommand{\bt}{\beta}
\newcommand{\ga}{\gamma}
\newcommand{\Ga}{\Gamma}
\newcommand{\de}{\delta}
\newcommand{\De}{\Delta}
\newcommand{\la}{\lambda}
\newcommand{\si}{\sigma}

\newcommand{\nn}{\nonumber}

\newcommand{\lqcd}{\Lambda_\mathrm{QCD}}
\newcommand{\cusp}{{\rm cusp}}

\newcommand{\GeV}{\mathrm{GeV}}

\newcommand{\Pythia}{\textsc{Pythia}\xspace}
\newcommand{\Herwig}{\textsc{Herwig}\xspace}

\allowdisplaybreaks[4]

\begin{document}

%%%%%%%%%%%%%%%%%%%%%%%%%%%%%%%%%%%%%%%%%%%%%%%%%%%%%%%%%%%%%%%%%%%%%%%%%%%%%%%%
% Title page
%%%%%%%%%%%%%%%%%%%%%%%%%%%%%%%%%%%%%%%%%%%%%%%%%%%%%%%%%%%%%%%%%%%%%%%%%%%%%%%%

\preprint{\vbox{
\hbox{DESY 17-111}
\hbox{NIKHEF 2017-031}
}}

\title{A case study of quark-gluon discrimination at NNLL$'$ in comparison to parton showers}

\author{Jonathan Mo}
\affiliation{Institute for Theoretical Physics Amsterdam and Delta Institute for Theoretical Physics, University of Amsterdam, Science Park 904, 1098 XH Amsterdam, The Netherlands}
\affiliation{Nikhef, Theory Group, Science Park 105, 1098 XG, Amsterdam, The Netherlands}

\author{Frank J.~Tackmann}
\affiliation{Theory Group, Deutsches Elektronen-Synchrotron (DESY), D-22607 Hamburg, Germany\vspace{0.5ex}}

\author{Wouter J.~Waalewijn}
\affiliation{Institute for Theoretical Physics Amsterdam and Delta Institute for Theoretical Physics, University of Amsterdam, Science Park 904, 1098 XH Amsterdam, The Netherlands}
\affiliation{Nikhef, Theory Group, Science Park 105, 1098 XG, Amsterdam, The Netherlands}

%%%%%%%%%%%%%%%%%%%%%%%%%%%%%%%%%%%%%%%%%%%%%%%%%%%%%%%%%%%%%%%%%%%%%%%%%%%%%%%%
\begin{abstract}
Predictions for our ability to distinguish quark and gluon jets vary by more than a factor of two between different
parton showers. We study this problem using analytic resummed predictions for the thrust event shape up to NNLL$'$
using $e^+e^- \to Z \to q \bar q$ and $e^+e^- \to H \to gg$ as proxies for quark and gluon jets.
We account for hadronization effects through a nonperturbative shape function, and include an estimate of both
perturbative and hadronization uncertainties. In contrast to previous studies, we find reasonable
agreement between our results and predictions from both \Pythia and \Herwig parton showers.
We find that this is due to a noticeable improvement in the description of gluon jets in the newest \Herwig 7.1
compared to previous versions.
\end{abstract}
%%%%%%%%%%%%%%%%%%%%%%%%%%%%%%%%%%%%%%%%%%%%%%%%%%%%%%%%%%%%%%%%%%%%%%%%%%%%%%%%

\maketitle

%%%%%%%%%%%%%%%%%%%%%%%%%%%%%%%%%%%%%%%%%%%%%%%%%%%%%%%%%%%%%%%%%%%%%%%%%%%%%%%%
\section{Introduction}
%%%%%%%%%%%%%%%%%%%%%%%%%%%%%%%%%%%%%%%%%%%%%%%%%%%%%%%%%%%%%%%%%%%%%%%%%%%%%%%%

The reliable discrimination between quark-initiated and gluon-initiated jets is a key goal of
jet substructure methods~\cite{Abdesselam:2010pt, Altheimer:2012mn, Altheimer:2013yza, Adams:2015hiv}.
It would provide a direct handle to distinguish hard processes that lead to the same number but
different types of jets in the final state.
A representative example is the search for new physics, where the signal processes typically produce quark jets,
while QCD backgrounds predominantly involve gluon jets from gluon radiation.

Jet substructure observables for quark-gluon discrimination have been studied extensively using both parton showers and analytic calculations~\cite{Gallicchio:2011xq,Gallicchio:2012ez,Larkoski:2013eya,Larkoski:2014pca,Bhattacherjee:2015psa,Badger:2016bpw,FerreiradeLima:2016gcz,Komiske:2016rsd,Davighi:2017hok,Gras:2017jty}. Much effort has been dedicated to identifying the most promising observables to achieve this goal. However, it has been known for a while that the discrimination power one obtains differs a lot between different parton shower predictions. A detailed study has been carried out in Refs.~\cite{Badger:2016bpw, Gras:2017jty}. It uses the classifier
%%%
\begin{align} \label{eq:classifier}
 \Delta = \frac12 \int\! \df \lambda\, \frac{\big[p_q(\la) - p_g(\la)\big]^2}{p_q(\la)+p_g(\la)}
\end{align}
%%%
to quantify the differences between the normalized quark and gluon distributions $p_{q,g}$ for an observable $\la$,
providing a measure of the quark-gluon separation.
The study found that the various parton showers agree well in their predictions for quark jets, which is not surprising since
much information on the shape of quark jets is available from LEP data. On the other hand, there is still very little
information on gluon jets available, and correspondingly the study identified the substantially different predictions for gluon jets as the main culprit.

Parton showers are formally only accurate to (next-to-)leading logarithmic order and do not provide an estimate of their intrinsic perturbative (resummation) uncertainties. Thus, it is not clear to what extent the observed differences are a reflection of (and thus consistent within) the inherent uncertainties, or whether only some of the parton showers obtain correct predictions.

In this paper, we address this issue by considering the thrust event shape for which we are able to obtain precise theoretical predictions from analytic higher-order resummed calculations, which can be used as a benchmark for parton-shower predictions.
An extensive survey of parton-shower predictions as carried out in Refs.~\cite{Badger:2016bpw, Gras:2017jty} is beyond our scope here. We will instead restrict ourselves to \Pythia~\cite{Sjostrand:2014zea} and \Herwig~\cite{Bellm:2017bvx}, as they represent the opposite extremes in the results of Refs.~\cite{Badger:2016bpw, Gras:2017jty}.

Thrust has been calculated to (next-to-)next-to-next-to-leading logarithmic ((N)NNLL) accuracy for quark jets produced in $e^+e^- \to q\bar q$ collisions~\cite{Becher:2008cf, Abbate:2010xh}. Here, we also obtain new predictions at NNLL$'$ for gluonic thrust using the toy process $e^+e^-\to H \to gg$, from which we can then calculate the quark-gluon classifier separation at NNLL$'$.\footnote{Since we consider normalized distributions, there is very little dependence on the specific hard processes we consider.}
Thrust is defined as 
%%%
\begin{equation}
  T = \max{}_{\hat t}\; \frac{\sum_i |\hat t \sdt {\vec p}_i|}{\sum_i |{\vec p}_i|}
  \,,\quad
  \tau = 1 - T
\,,\end{equation}
%%%
where the sum over $i$ runs over all final-state particles. For $\tau \ll 1$, the final state consists of two back-to-back jets. The radiation in these jets is probed by $\tau$, since in this limit
%%%
\begin{align}
\tau = \frac{M_1^2 + M_2^2}{Q^2}
\,,\end{align}
%%%
where $M_{1,2}$ are the invariant masses of the two (hemisphere) jets and $Q$ is the invariant mass of the collision.
Thrust corresponds closely to the generalized angularity $(\kappa,\beta) = (1,2)$, which was one of the benchmark observables
considered in Refs.~\cite{Badger:2016bpw,Gras:2017jty}. (The difference is that for the latter one only sums over particles within a certain jet radius around the thrust axis).

Our numerical results include resummation up to NNLL$'$ resummation and include nonperturbative hadronization corrections through a shape function~\cite{Korchemsky:1999kt, Korchemsky:2000kp, Hoang:2007vb, Ligeti:2008ac}. We assess the perturbative uncertainty through appropriate variations of the profile scales~\cite{Ligeti:2008ac, Abbate:2010xh}, and the nonperturbative uncertainty by varying the nonperturbative parameter $\Omega$, which quantifies the leading nonperturbative corrections.

\begin{figure}[t]
\centering
\includegraphics[width=\columnwidth]{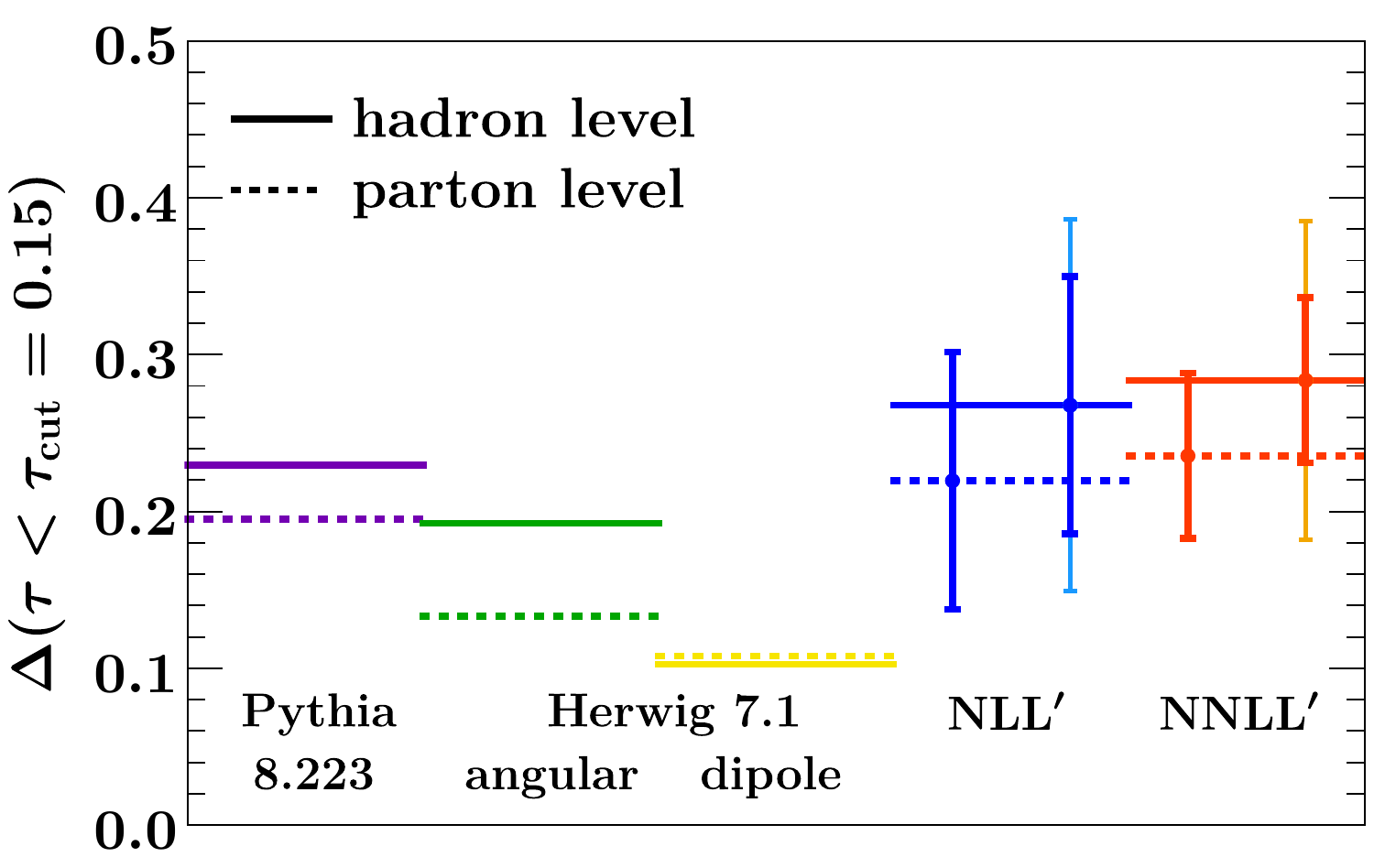}
\caption{The quark-gluon classifier separation $\De$ for $\tau < 0.15$ from \Pythia 8.223 (violet), \Herwig 7.1 angular-ordered shower (green) and dipole shower (yellow) compared to analytic resummation at NLL$'$ (blue) and NNLL$'$ (red). The results at parton and hadron level are shown in dotted and solid, respectively. The uncertainty bars on the resummed results show the perturbative uncertainty and also the sum of perturbative and hadronization uncertainties (lighter outer bars at hadron level). \vspace{-1ex}}
\label{fig:moneyplot}
\end{figure}

\fig{moneyplot} shows the classifier separation for quark-gluon discrimination in \eq{classifier} at parton and hadron level obtained from our analytic predictions, compared to \Pythia 8.223~\cite{Sjostrand:2014zea} and \Herwig 7.1~\cite{Bellm:2017bvx}. Our resummed results are shown at NLL$'$ and NNLL$'$, and include an estimate of the perturbative and hadronization uncertainty. As we do not combine our NNLL$'$ prediction with the full fixed-order NNLO result, which would become relevant at large $\tau$, we restrict the integration range here to $\tau < 0.15$. Both \Pythia's parton shower and \Herwig's default angular-ordered shower are consistent with our results.
We observe that the tension between these two showers is much reduced here compared to what was found in Refs.~\cite{Badger:2016bpw,Gras:2017jty}. As we will see later, this is due to an improved description of gluon jets in \Herwig 7.1 compared to earlier versions. Specifically, the parton shower now preserves the virtuality rather than the transverse momentum after multiple emissions, and has been tuned to gluon data for the first time~\cite{Richardson}.
For comparison, we also include results obtained using \Herwig's dipole shower, which still gives substantially lower
predictions compared to the others.

The outline of this paper is as follows: In \sec{calc} we present the details of our calculation. Many of the ingredients can be found in the literature but are reproduced here (and in appendices) to make the paper self-contained. We present numerical results in \sec{res} for the thrust distribution of quark and gluons jets, as well as the classifier separation calculated from it, and performing comparisons to \Pythia and \Herwig. In \sec{conc} we conclude.

%%%%%%%%%%%%%%%%%%%%%%%%%%%%%%%%%%%%%%%%%%%%%%%%%%%%%%%%%%%%%%%%%%%%%%%%%%%%%%%%
\section{Calculation}
\label{sec:calc}
%%%%%%%%%%%%%%%%%%%%%%%%%%%%%%%%%%%%%%%%%%%%%%%%%%%%%%%%%%%%%%%%%%%%%%%%%%%%%%%%

The cross section for thrust factorizes~\cite{Catani:1992ua,Korchemsky:1999kt,Fleming:2007qr,Schwartz:2007ib} 
%%%
\begin{align} \label{eq:fact}
   \frac{\df \si_i}{\df \tau} &= \si_{i,0}\, |C_i(Q,\mu)|^2 \int\! \df s_1\, J_i(s_1,\mu) \int\! \df s_2\, J_i(s_2,\mu) 
   \\ & \quad \times
   \int\! \df k\, S_i(k,\mu)\,
   \de\Big(\tau - \frac{s_1 + s_2}{Q^2} - \frac{k}{Q}\Big)
   +  \frac{\df \si_i^{\rm nons}}{\df \tau}
\,,\nn\end{align}
%%%
where the label $i=q$ corresponds to the hard process $Z \to q \bar q$ and $i=g$ corresponds to $H \to gg$. The Born cross section is denoted by $\si_{i,0}$, with hard virtual corrections contained in the hard Wilson coefficient $C_i$. The jet functions $J_i$ describes the invariant masses $s_{1,2}$ of the energetic (collinear) radiation in the jets. The soft function $S_i$ encodes the contribution $k$ of soft radiation to the thrust measurement. Contributions that do not factorize in this manner are suppressed by relative $\ord{\tau}$ and are contained in the nonsingular cross section $\df \si_i^{\rm nons}/\df \tau$.

%===============================================================================
\subsection{Resummation}
%===============================================================================

For $\tau \ll 1$ the thrust spectrum contains large logarithms of $\tau$, that we resum by utilizing the renormalization group evolution that follows from the factorization in \eq{fact}. This is accomplished by evaluating $C_i$, $J_i$, and $S_i$ at their natural scales
 %%%
 \begin{align} \label{eq:canonical}
   \mu_C \simeq Q
   \,, \quad
   \mu_J \simeq \sqrt{\tau Q}
   \,, \quad
   \mu_S \simeq \tau Q   
 \,,\end{align}
 %%%
where they each do not contain large logarithms, and evolving them to a common (and arbitrary) scale $\mu$.
The precise resummation scales and their variations used in our numerical results are given in \eq{profile}.
 
The renormalization group equations of the hard, jet, and soft functions are given by
%%%
\begin{align} \label{eq:RGE}
\mu \frac{\df}{\df\mu}\, C_i(Q, \mu) &= \gamma_C^i(Q, \mu)\, C_i(Q, \mu)
\,, \\
\gamma_C^i(Q, \mu) &=
\Gamma_\cusp^i[\alpha_s(\mu)] \ln\frac{Q^2}{\mu^2} + 2\gamma_C^i[\alpha_s(\mu)]
\,,\nn \\
\mu \frac{\df}{\df \mu}\, J_i(s, \mu) &= \int\! \df s'\, \gamma_J^i(s-s',\mu)\, J_i(s', \mu)
\,,\nn\\
\gamma_J^i(s, \mu)
&= -2 \Gamma^i_{\cusp}[\alpha_s(\mu)]\,\frac{1}{\mu^2} \biggl[\frac{\mu^2}{s}\biggr]_+ \!\!\!+ \gamma_J^i[\alpha_s(\mu)] \delta(s)
\,,\nn \\
\mu\frac{\df}{\df\mu}\, S_i(k, \mu)
&= \int\! \df k'\, \gamma_S^i(k\! - k', \mu)\, S_i(k', \mu)
\,,\nn\\
\gamma_S^i(k, \mu)
&= 4\,\Gamma_\cusp^i[\alpha_s(\mu)]\, \frac{1}{\mu} \biggl[\frac{\mu}{k}\biggr]_+ \!\!+
\gamma_S^i[\alpha_s(\mu)]\, \delta(k)
\,,\nn \end{align}
%%%
and involve the cusp anomalous dimension $\Gamma_\cusp^i(\al_s)$~\cite{Korchemsky:1987wg} and a noncusp term $\ga_{C,J,S}^i(\al_s)$. (The factor of 2 in front of $\ga_C^i(\al_s)$ is included to be consistent with our conventions in e.g.~Ref.~\cite{Moult:2015aoa}.) The $\mu$ independence of the cross section in \eq{fact} implies the consistency condition
%%%
\begin{align} \label{eq:consistency}
4\gamma_C^i(\alpha_s) + 2\gamma_J^i(\alpha_s) + \gamma_S^i(\alpha_s) &= 0
\,.\end{align}
%%%
We employ analytic solutions to the RG equations, which for the jet and soft function follow from Refs.~\cite{Balzereit:1998yf, Neubert:2004dd, Fleming:2007xt}. For our implementation we use the results for the RG solution and plus-function algebra derived in Ref.~\cite{Ligeti:2008ac}.

%%%
\begin{table}[t!]
  \centering
  \begin{tabular}{l | c c c c c c}
  \hline \hline
  & $C_i, J_i, S_i$ & $\gamma_C^i, \gamma_J^i, \gamma_S^i$ & $\Gamma_{\rm cusp}, \beta$ \\ \hline
  LL & $0$-loop & - & $1$-loop \\
  NLL & $0$-loop & $1$-loop & $2$-loop \\
  NLL$'$ & $1$-loop & $1$-loop & $2$-loop \\
  NNLL & $1$-loop & $2$-loop & $3$-loop \\
  NNLL$'$ & $2$-loop & $2$-loop & $3$-loop \\
  NNNLL & $2$-loop & $3$-loop & $4$-loop \\
  \hline\hline
  \end{tabular}
  \caption{Perturbative ingredients at different orders in resummed perturbation theory.}
\label{tab:orders}
\end{table}
%%%

The ingredients that enter the cross section at various orders of resummed perturbation theory are summarized in Table~\ref{tab:orders}. Our best predictions are at NNLL$'$ order, which is closer to NNNLL than NNLL, as the inclusion of the two-loop fixed-order ingredients has a larger effect than the three-loop non-cusp and four-loop cusp anomalous dimension. Our NNLL$'$ predictions require the two-loop hard function~\cite{Kramer:1986sg,Matsuura:1987wt,Matsuura:1988sm,Becher:2006mr,Idilbi:2006dg,Harlander:2009bw,Pak:2009bx,Berger:2010xi}, jet function~\cite{Becher:2006qw,Becher:2010pd}, and soft function~\cite{Kelley:2011ng,Hornig:2011iu}. The RG evolution involves the three-loop QCD beta function~\cite{Tarasov:1980au,Larin:1993tp}, three-loop cusp anomalous dimension~\cite{Moch:2004pa} and two-loop non-cusp anomalous dimensions~\cite{Idilbi:2006dg,Becher:2006mr,Idilbi:2006dg,Becher:2009th}. All necessary expressions are collected in the appendices. In our numerical analysis we take $\al_s(m_Z) = 0.118$.

%===============================================================================
\subsection{Nonsingular Corrections}
%===============================================================================

To obtain a reliable description of the thrust spectrum for large values of $\tau$ we also need to include the nonsingular $\df \si_i^{\rm nons}/\df \tau$ in \eq{fact}. These are obtained from the full $\ord{\al_s}$ expressions
%%%
\begin{align} \label{eq:nons}
  \frac{\df \si_q}{\df \tau} &= \si_{q,0}\, \frac{\al_s C_F}{2\pi}\, \frac{1}{\tau (\tau-1)}\biggl[3 - 9 \tau - 3 \tau^2 + 9 \tau^3
  \nn \\ & \quad
  - (4-6\tau+6\tau^2) \ln\frac{1- 2\tau}{\tau} \biggr]
  \,, \nn \\
\frac{\df \si_g}{\df \tau} &= \si_{g,0}\,\frac{\al_s}{2\pi} \biggl\{C_A\, \frac{1}{3\tau(\tau\!-\!1)}\biggl[11 \!-\! 68\tau \!+\! 144\tau^2 \!-\! 132\tau^3
    \nn \\ & \quad
     +45\tau^4 - 12(1 - 2\tau + 3\tau^2 - 2\tau^3 + \tau^4)\ln\frac{1-2\tau}{\tau} \biggr]
  \nn \\ & \quad
    +  T_F n_f\, \frac{2}{3\tau}\biggl[2 - 21\tau + 60\tau^2 -45\tau^3
      \nn \\ & \quad
      + 6\tau(1 - 2\tau + 2\tau^2)\ln\frac{1-2\tau}{\tau} \biggr] \biggr\}
,\end{align}
%%%
and subtracting the terms that are singular in the $\tau\to 0$ limit, which are contained in the NLL$'$ resummed result.
Adding the $\ord{\alpha_s}$ nonsingular corrections to the NLL$'$ resummed cross section then yields the final
matched NLL$'+$NLO result.
The above result for the quark case has been known for a long time~\cite{Ellis:1980wv}. The gluon result was obtained by squaring and summing the helicity amplitudes in Ref.~\cite{Schmidt:1997wr} and performing the required phase-space integrations to project onto the $\tau$ spectrum. At NNLL$'$ we would also need the full $\ord{\al_s^2}$ terms to obtain the matched NNLL$'+$NNLO result, so we restrict ourselves to small $\tau < 0.15$ in this case, such that we can neglect the nonsingular corrections.

%===============================================================================
\subsection{Hadronization Effects}
%===============================================================================

The soft function in the factorization theorem in \eq{fact} accounts for both perturbative
soft radiation and nonperturbative hadronization effects. The hadronization effects can
be taken into account by factorizing the full soft function as~\cite{Korchemsky:1999kt, Hoang:2007vb, Ligeti:2008ac}
%%%
\begin{align} \label{eq:nonp}
S_i(k,\mu) = \int\! \df k'\, S_i^{\rm pert}(k-k',\mu) F_i(k')
\,,\end{align}
%%%
where $S_i^{\rm pert}(k,\mu)$ contains the perturbative corrections and
$F_i(k)$ is a nonperturbative shape function encoding hadronization effects.
This treatment is known to provide an excellent description of hadronization
effects in $B$-meson decays~\cite{Bernlochner:2013gla} and $e^+e^-$ event shapes~\cite{Abbate:2010xh}.
It has furthermore been successfully utilized for quark and gluon jet mass spectra
in hadron collisions~\cite{Stewart:2014nna}.

The shape function $F_i(k)$ is normalized to unity and has typical support for
$k \sim \lqcd$. It should vanish at $k = 0$ and fall off exponentially for $k\to \infty$.
We use a simple ansatz that satisfies these basic criteria~\cite{Stewart:2014nna}
%%%
\begin{align} \label{eq:shape}
F_i(k') = \frac{k'}{\Omega_i^2} e^{-k'/ \Omega_i}
\,.\end{align}
%%%
The parameter $\Omega_i$ captures the leading nonperturbative correction in the tail of the distribution, where it leads to a shift $\tau \to \tau + 2\Omega_i/Q$. We take $\Omega_q = 0.4$~\cite{Abbate:2010xh} and assume Casimir scaling, $\Omega_g = \Omega_q C_A/C_F$. As an estimate of the nonperturbative uncertainty we vary $\Omega_q$ and $\Omega_g$ over a large range as discussed above \eq{nonpunc}. In the peak of the distribution in principle the full functional form of $F_i(k)$ enters. However, given the large uncertainties for $\Omega_i$ we currently include, the precise functional form of $F_i$ is not yet of practical importance.

%===============================================================================
\subsection{Estimation of Uncertainties}
\label{sec:uncertainties}
%===============================================================================

The canonical scales in \eq{canonical} do not properly take into account the transition from the resummation region into the fixed-order region where $\tau$ is no longer small, or into the nonperturbative region for $\tau \lesssim \lqcd/Q$. A smooth transition between these different regimes is accomplished using profile scales~\cite{Ligeti:2008ac, Abbate:2010xh}.

For the choice of profiles scales and the estimation of perturbative uncertainties
through their variations we follow the approach of Ref.~\cite{Stewart:2013faa}
adapted to the thrust-like resummation as in Ref.~\cite{Gangal:2014qda}.
The central values for the profile scales are taken as
%%%
\begin{align} \label{eq:profile}
& \mu_H = \mu
 \,, \quad
 \mu_S(\tau) = \mu  f_{\rm run}(\tau)
 \,, \quad
 \mu_J(\tau) = \sqrt{\mu_S(\tau) \mu}  
\,, \nn \\
&f_{\rm run} (\tau) = \left\{
\begin{tabular}{ll}
$\tau_0(1+ \frac{\tau^2}{(2\tau_0)^2})$ & $\tau\leq 2 \tau_0$
\\
$\tau$ & $2\tau_0 \leq \tau \leq \tau_1 $
\\
$\tau + \frac{(2-\tau_2-\tau_3)(\tau-\tau_1)^2}{2(\tau_2-\tau_1)(\tau_3-\tau_1)}$ & $\tau_1 \leq \tau \leq \tau_2$
\\
$1 - \frac{(2-\tau_1-\tau_2)(\tau-\tau_3)^2}{2(\tau_3-\tau_1)(\tau_3-\tau_2)}$ & $\tau_2 \leq \tau \leq \tau_3$
\\
$1$ & $\tau_3 \leq \tau$
\end{tabular}
\right.
\end{align}
%%%
Here, $\tau_0$ determines the boundary between the resummation and nonperturbative region, where the jet and soft scales approach $\sqrt{\tau_0}Q$ and $\tau_0 Q$ respectively. We choose $\tau_0 = 3 \GeV / Q$, so that $\mu_J$, $\mu_S$ are always greater than $\Lambda_{\mathrm{QCD}}$. From $\tau_0$ onwards we have the canonical resummation scales in \eq{canonical} up to $\tau_1 = 0.1$, where the different scales are still well separated. Then we smoothly turn the resummation off by letting $f_{\mathrm{run}}(\tau)$ go to 1. The resummation is completely turned off at $\tau_3 = 1/3$, where the singular and nonsingular contributions start to cancel each other exactly at $\ord{\alpha_s}$. The central curve of our prediction corresponds to
%%%
\begin{align}
  \mu = Q\,, \quad \tau_0 = \frac{3 \GeV}{Q}\,, \quad \tau_1 = 0.1
  \,, \nn \\
   \quad \tau_2 = \frac{\tau_1 + \tau_3}{2}\,, \quad \tau_3 = \frac{1}{3}
\,.\end{align}
%%%

The perturbative uncertainty is obtained as the quadratic sum of a fixed-order and a resummation contribution,
%%%
\begin{align} \label{eq:pertunc}
\delta_{\rm pert} = \sqrt{\delta_{\rm FO}^2 + \delta_{\rm resum}^2}
\,.\end{align}
%%%
The fixed-order uncertainty is estimated by the maximum observed deviation from varying the parameter $\mu$
in \eq{profile} by a factor of two,
%%%
\begin{align}
  \de_{\rm FO} (\tau) = \max_{\mu = \{2Q, Q/2\}} \bigg|\frac{\df \si}{\df \tau} - \frac{\df \si_{\rm central}}{\df \tau} \bigg|
\,.\end{align}
%%%
The resummation uncertainty is estimated by varying $\mu_{J,S}$ by~\cite{Gangal:2014qda}
%%%
\begin{align}
  \mu^{\rm vary}_S(\tau,\al) &= f_{\rm vary}^\al(\tau)\, \mu_S(\tau)
  \,,  \\
  \mu^{\rm vary}_J(\tau,\al,\bt) &= \mu^{\rm vary}_S(\tau,\al)^{1/2-\bt} \mu^{1/2+\bt} 
  \,,\nn \\  
  f_{\rm vary}(\tau) &= \left\{
\begin{tabular}{ll}
$2(1-\tau^2/\tau_3^2)$ & $\tau\leq \tau_3/2$
\\
$1 + 2(1-\tau/\tau_3)^2$ & $\tau_3/2 \leq \tau \leq \tau_3$
\\
$1$ & $\tau_3 \leq \tau$
\end{tabular}
\right.  
\nn \end{align}
%%%
and taking the maximum absolute deviation among all variations
%%%
\begin{align}
  \de_{\rm resum} (\tau) = \max_{(\al,\bt)} \bigg|\frac{\df \si}{\df \tau} - \frac{\df \si_{\rm central}}{\df \tau} \bigg|
\,,\end{align}
%%%
with $(\al,\bt) \in \{(1,0),(-1,0),(0,1/6), (0,- 1/6)\}$. Furthermore, we vary the transition points $\tau_0$ and $\tau_1$ of the resummation region by $\pm 25\%$. These variations however have a much smaller effect than the $\al,\bt$ variations, and their effect on the final resummation uncertainty is almost negligible.

To account for hadronization uncertainties, we separately vary $\Omega_q$ by $\pm 50\%$, $\Omega_g$ by $\pm 50\%$, and simultaneously vary $\Omega_q$ and $\Omega_g$ by $\pm 75\%$. The hadronization uncertainty $\de_{\rm nonp}$ is then taken as the maximum deviation under these variations. It is treated as a separate uncertainty source uncorrelated from the perturbative uncertainty, with the total uncertainty given by their quadratic sum,
%%%
\begin{align} \label{eq:nonpunc}
\delta = \sqrt{\delta_{\rm pert}^2 + \delta_{\rm nonp}^2}
.\end{align}
%%%

We follow a similar procedure to assess the uncertainty on the classifier separation. However, we do not vary the quark distribution and gluon distribution simultaneously, as varying them in opposite directions would lead to an unrealistic inflation of the uncertainty. Instead, we obtain the uncertainty on the classifier separation by taking the central quark result and varying the gluon distribution, and vice versa. This amounts to treating the perturbative uncertainties in the quark and gluon distributions as uncorrelated sources of uncertainties.

%%%%%%%%%%%%%%%%%%%%%%%%%%%%%%%%%%%%%%%%%%%%%%%%%%%%%%%%%%%%%%%%%%%%%%%%%%%%%%%%
\section{Results}
\label{sec:res}
%%%%%%%%%%%%%%%%%%%%%%%%%%%%%%%%%%%%%%%%%%%%%%%%%%%%%%%%%%%%%%%%%%%%%%%%%%%%%%%%

\begin{figure*}[t]
\includegraphics[width=\columnwidth]{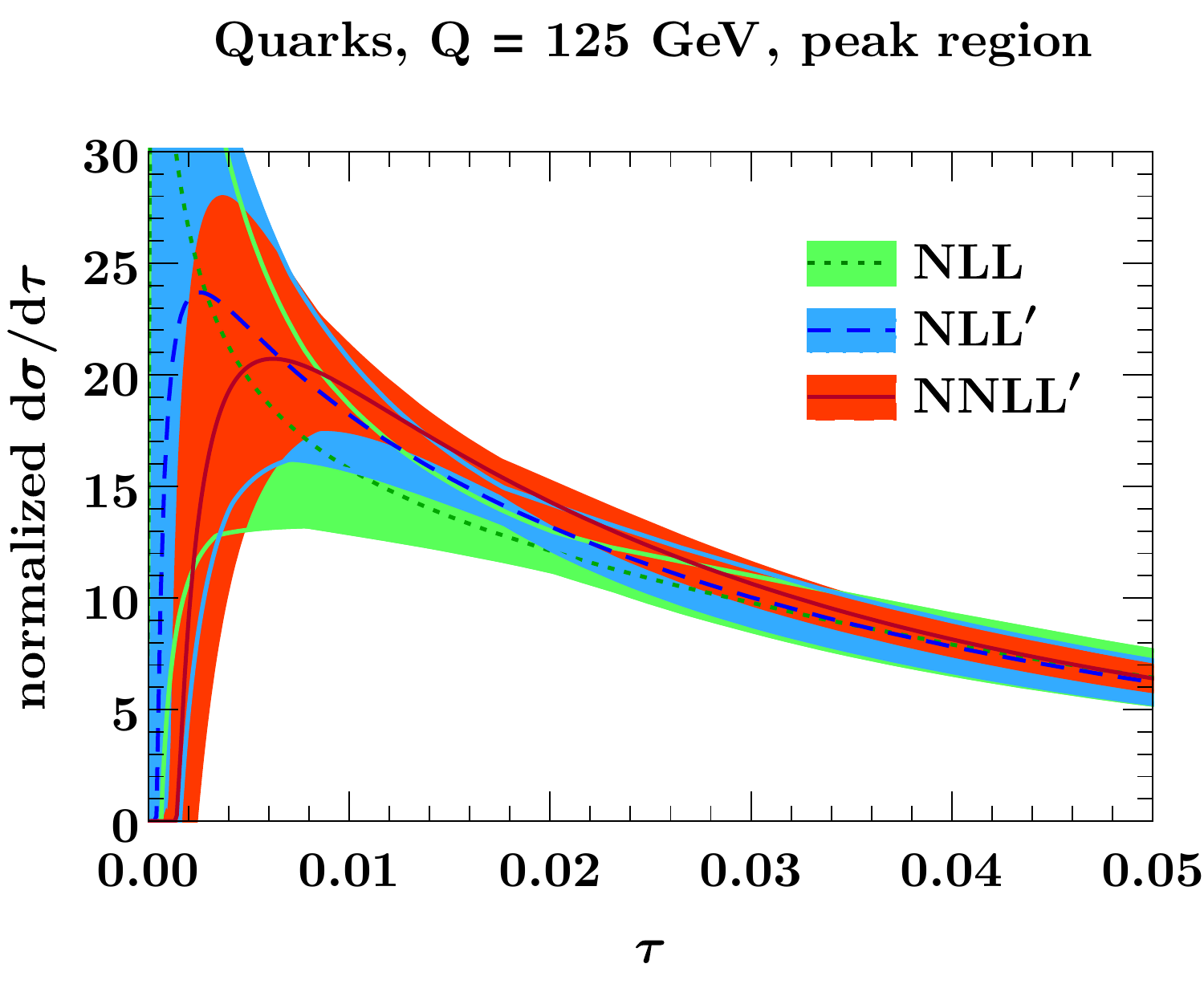}%
\hfill%
\includegraphics[width=\columnwidth]{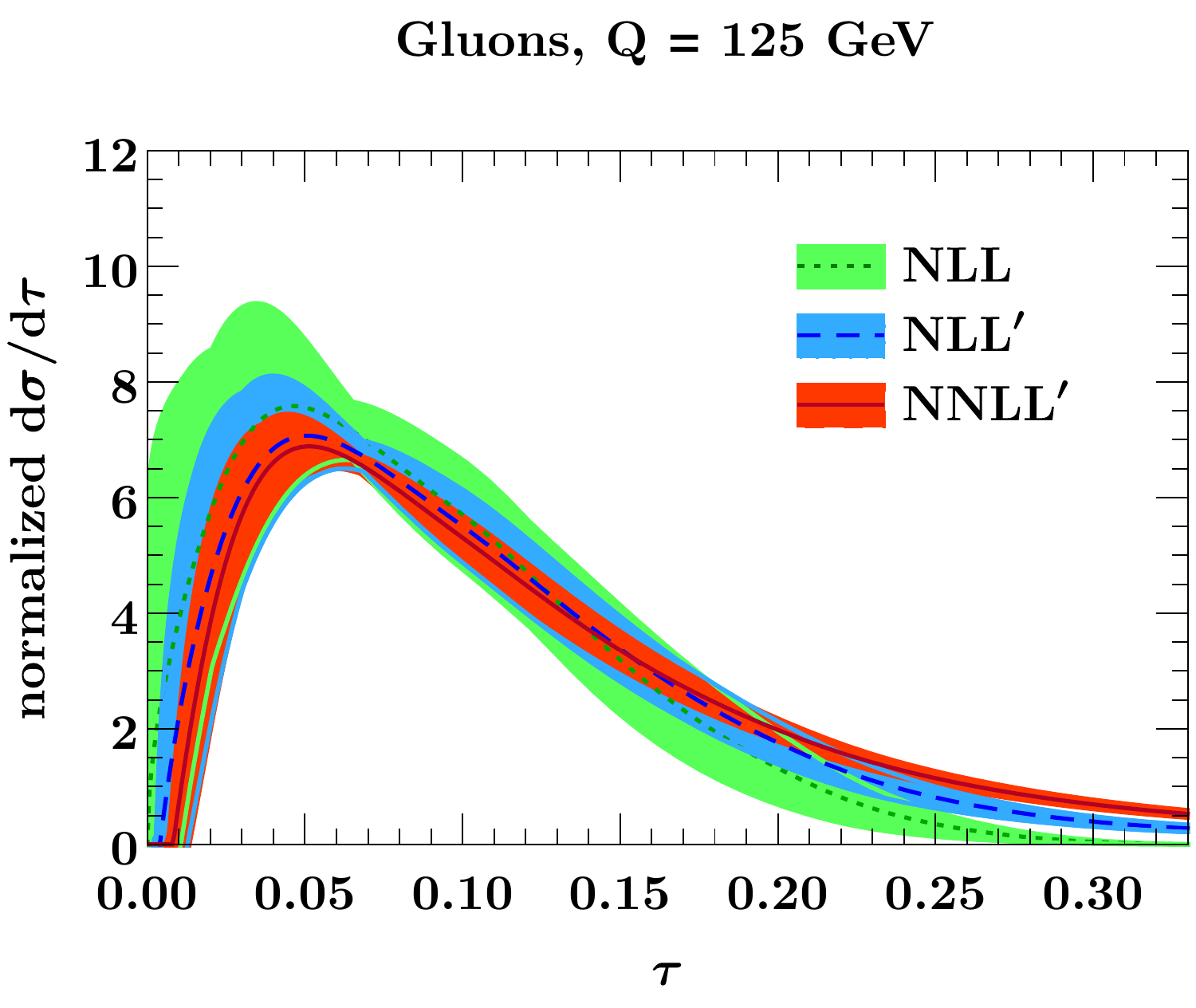}
\caption{The normalized thrust spectrum at NLL (green), NLL$'$ (blue), and NNLL$'$ (orange) for quarks (left panel) and gluons (right panel). Since the quark distribution on the left is peaked at small $\tau$, we restrict the plot range to $\tau<0.05$. The bands indicate the perturbative uncertainty at each order, obtained using \eq{pertunc}.}
\label{fig:orders}
\end{figure*}

\begin{figure*}[t]
\includegraphics[width=\columnwidth]{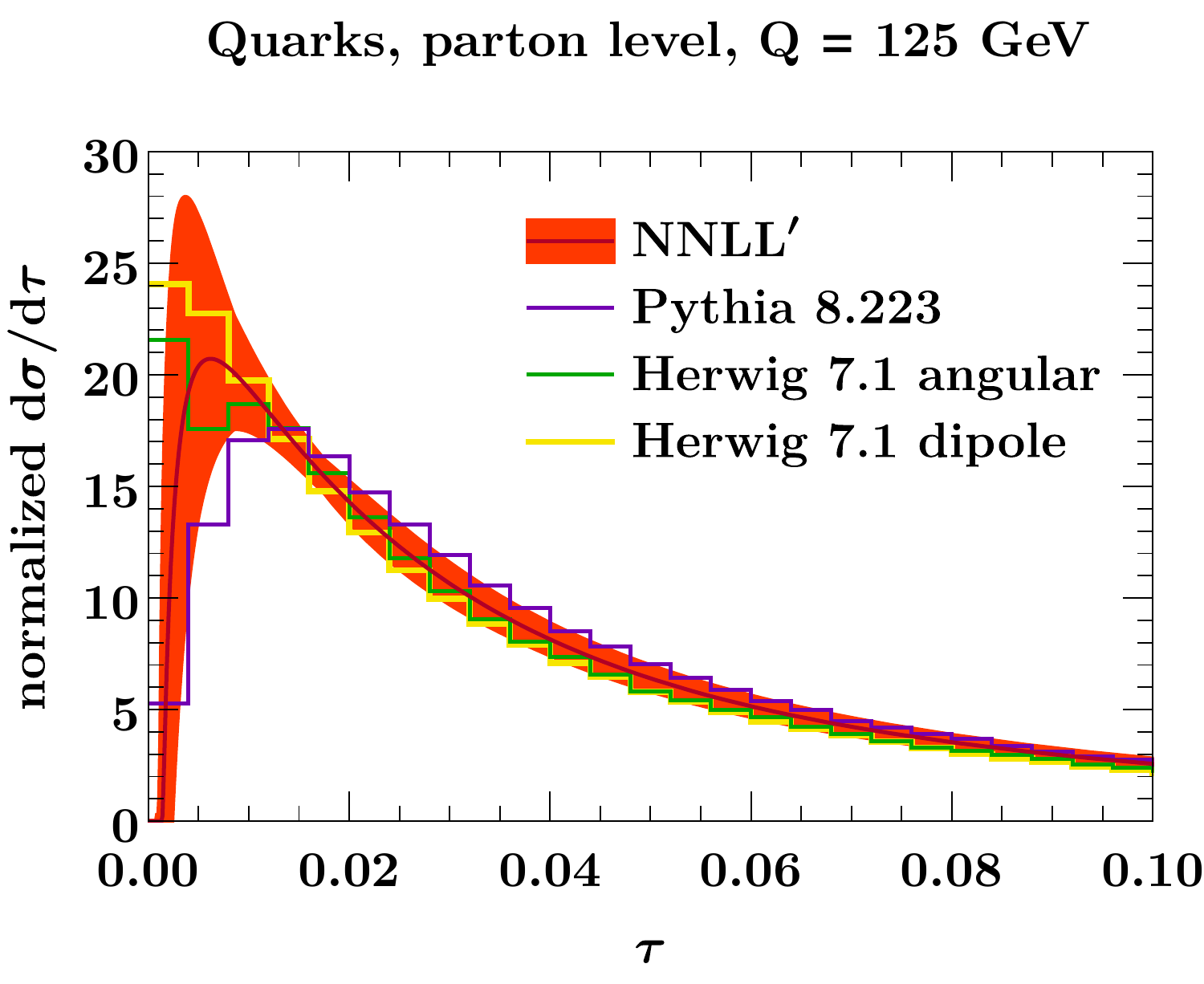}%
\hfill%
\includegraphics[width=\columnwidth]{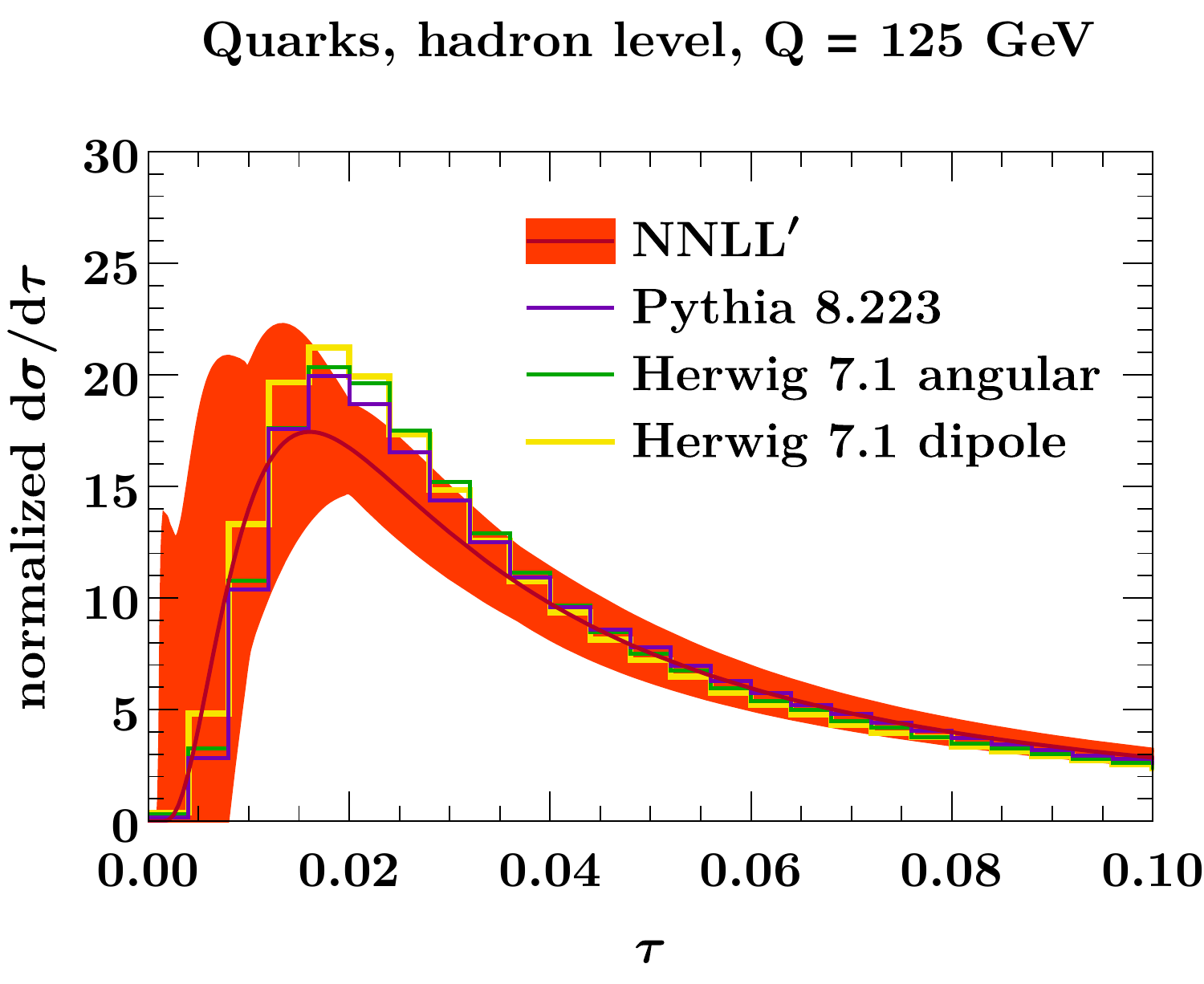}
\caption{The normalized thrust spectrum for quarks at NNLL$'$ (orange band) compared to \Pythia (violet) and \Herwig's angular-ordered (green) and dipole shower (yellow) at parton level (left panel) and hadron level (right panel). The band in the left panel shows the perturbative uncertainty in \eq{pertunc}. In the right panel, it shows the sum of perturbative and nonperturbative uncertainties as in \eq{nonpunc}.}
\label{fig:quark}
\end{figure*}

\begin{figure*}[t]
\includegraphics[width=\columnwidth]{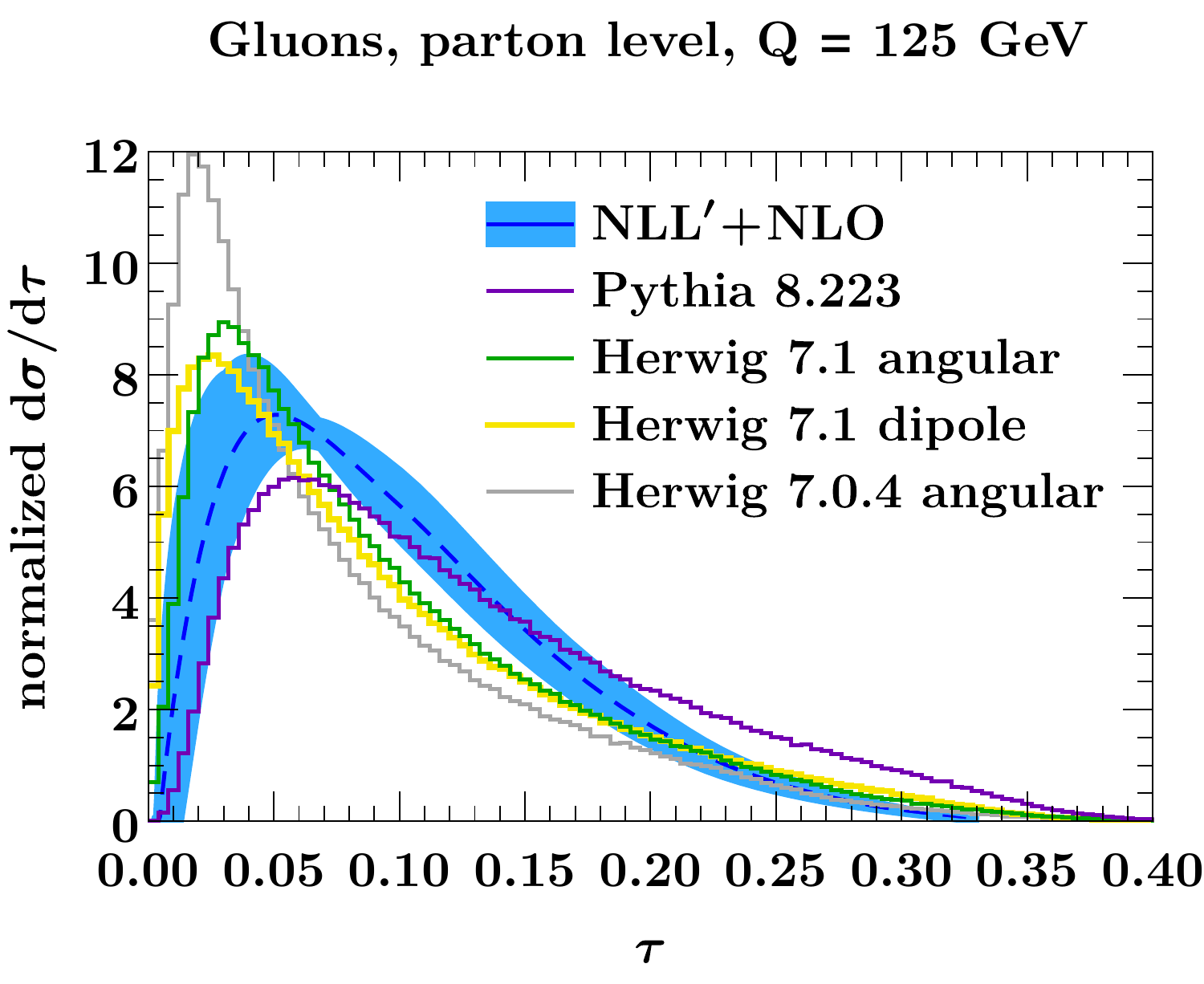}%
\hfill%
\includegraphics[width=\columnwidth]{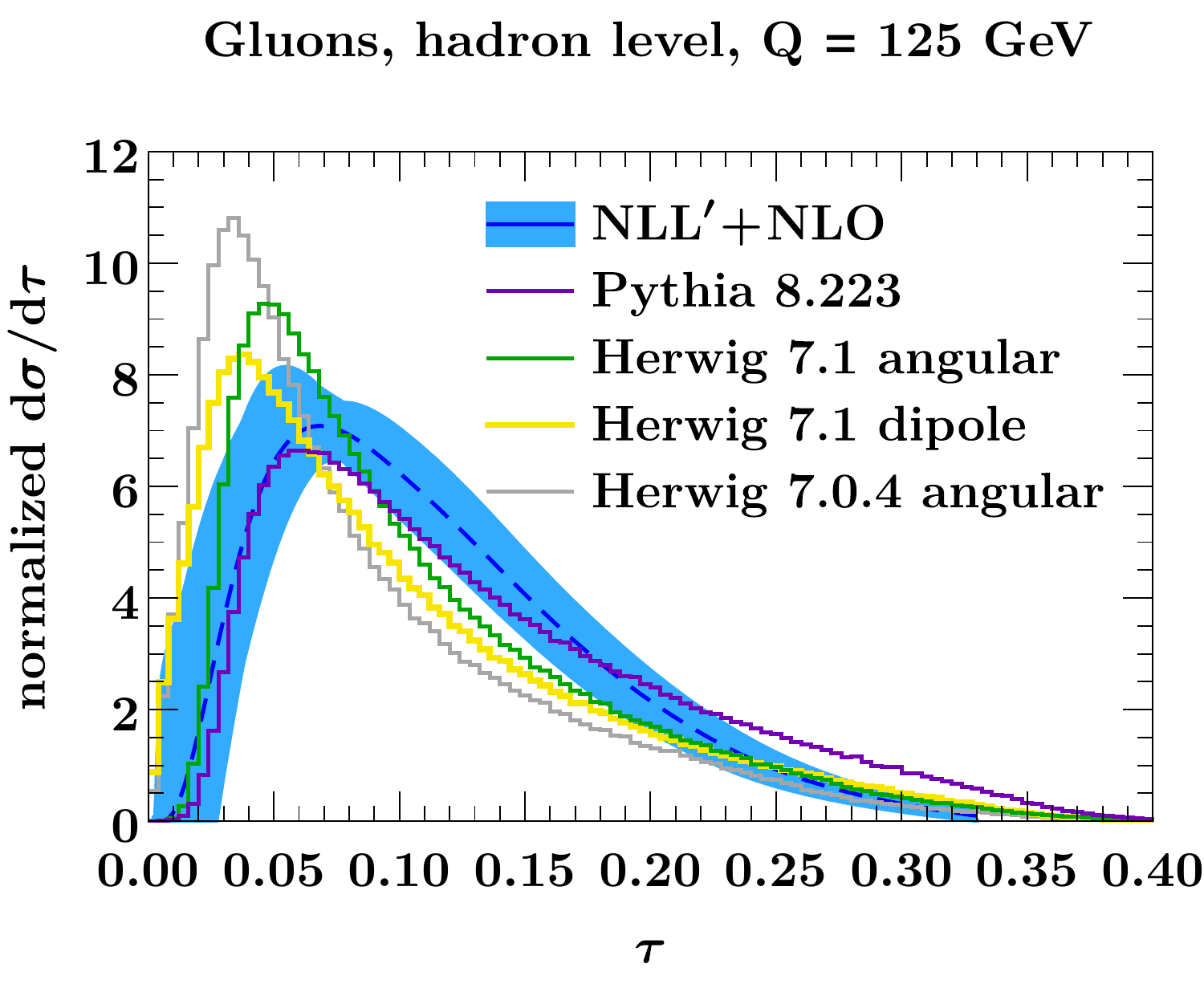}
\caption{The normalized thrust spectrum for gluons at NLL$'+$NLO (blue band) compared to \Pythia (violet) and \Herwig's angular-ordered (green) and dipole shower (yellow) at parton level (left panel) and hadron level (right panel). The band in the left panel shows the perturbative uncertainty in \eq{pertunc}. In the right panel, it shows the sum of perturbative and nonperturbative uncertainties as in \eq{nonpunc}. The result from the angular-ordered shower in \Herwig 7.0.4 is shown in light gray, which differs significantly from the resummed results, highlighting the noticeable improvement in \Herwig 7.1. \vspace{-1ex}}
\label{fig:gluon}
\end{figure*}

\begin{figure}[t]
\centering
\includegraphics[width=\columnwidth]{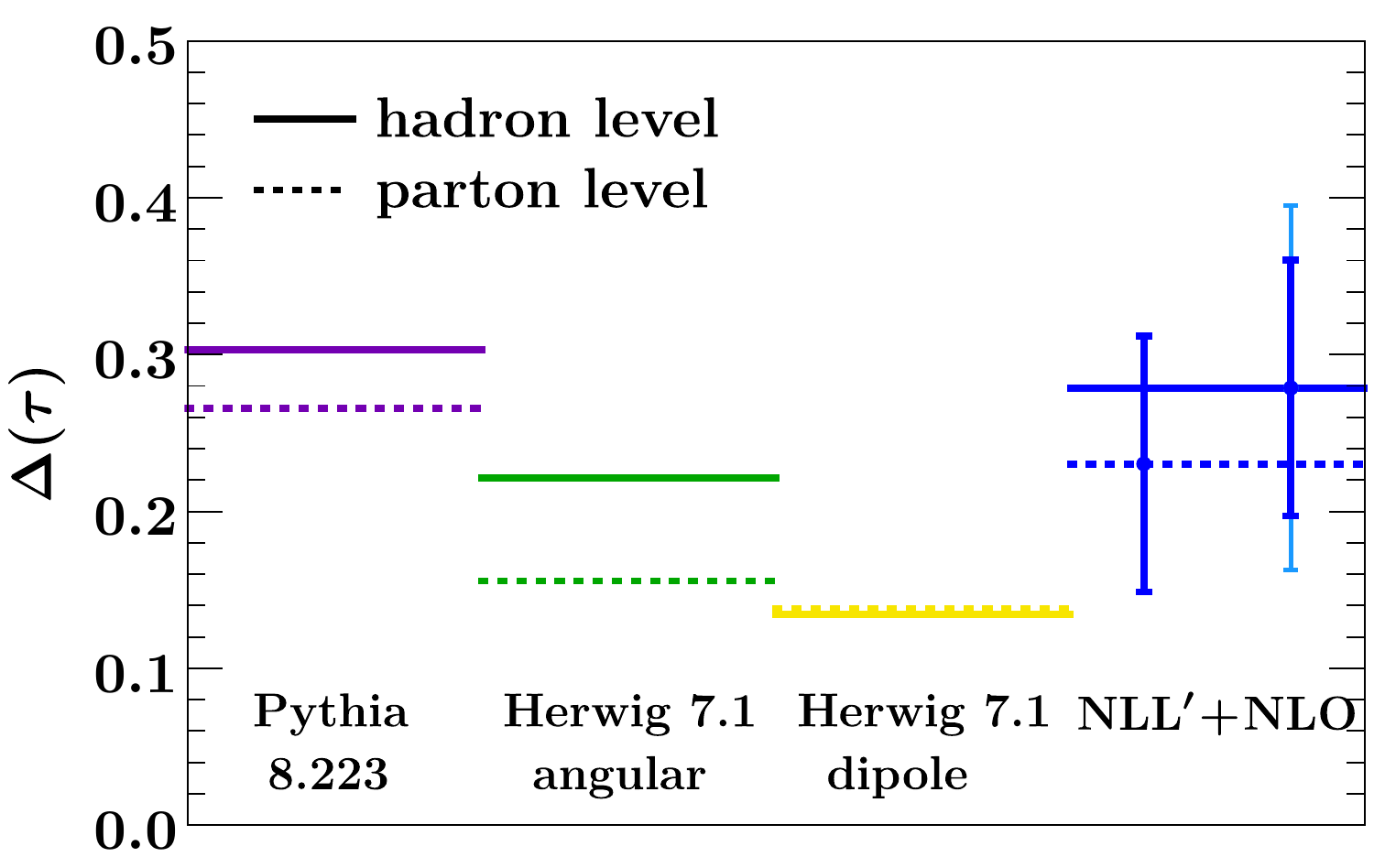}
\caption{Analogous to \fig{moneyplot} but without a cut on $\tau$. \vspace{-2ex}}
\label{fig:classifier}
\end{figure}

We now present our numerical results and compare these to \Pythia and \Herwig. We restrict ourselves to normalized distributions, as these are the input entering in the classifier separation in \eq{classifier}.

\fig{orders} shows the thrust spectrum for quarks and gluons at various orders in resummed perturbation theory. The bands show the perturbative uncertainty, obtained using the procedure described in \sec{uncertainties}. The overlapping uncertainty bands suggest that our uncertainty estimate is reasonable, and the reduction of the uncertainty  at higher orders indicates the convergence of our resummed predictions. This is not true for large values of $\tau$, because we did not include the nonsingular corrections $\df \si^{\rm nons}/\df \tau$ that are important in this region.

In \figs{quark}{gluon} we compare our predictions for quarks and gluons at parton and hadron level to \Pythia and \Herwig.
Note that the peak of the quark distribution is in the nonperturbative regime $\tau Q \lesssim \lqcd$. Therefore we restrict to $\tau < 0.1$ when considering the quark distributions in \fig{quark}, allowing the use of the NNLL$'$ result. On the other hand, the gluon distribution peaks at much higher values, and so we consider the gluon distribution over the full $\tau$ range using the matched result at NLL$'$+NLO.

For quarks at parton level, shown in the left panel of \fig{quark}, both \Pythia and \Herwig agree well with the resummed result and also with each other. The only exception is in the nonperturbative regime at very small $\tau$, where the comparison of parton-level predictions is not very meaningful. At the hadron level (right panel of \fig{quark}) we also include the nonperturative uncertainty in our band, and our predictions agree well with \Pythia and \Herwig. Note that \Pythia and \Herwig at hadron level agree with each other even better than at parton level. This is of course not surprising, as their hadronization models have been tuned to the same LEP data. The differences seen at parton level are likely due to a higher shower cutoff scale in \Herwig (which would also explain the events with $\tau=0$), and is compensated for by the hadronization~\cite{Gras:2017jty}.

We now turn to the results for gluons shown in \fig{gluon}. Here, there differences between \Pythia and \Herwig are much larger at both parton and hadron level. At parton level and small values of $\tau$, the \Herwig 7.1 and \Pythia predictions touch opposite sides of the uncertainty band of the NLL$'$+NLO result. Thus, although the differences in the parton shower results are clearly sizeable, they might still be considered to be within their intrinsic uncertainties, also since the formal accuracy of the showers is less than that of the NLL$'+$NLO result.
For large values beyond $\tau > 0.2$ there are differences between \Pythia and our result. However, this region is not described by the resummation but the fixed-order calculation. At NLO there are only three partons, so $\tau \leq 1/3$. Although \Pythia produces events with $\tau>1/3$, it does not do so with any formal accuracy, since the parton shower is built from collinear/soft limits of QCD which do not apply here.

For gluons at hadron level, \Pythia agrees well with our result. The agreement for \Herwig 7.1 is less good, though the differences are not that large either. However, we see that the angular ordered shower from \Herwig 7.0.4 shown by the gray lines shows clear discrepancies from our predictions. (It also yields similarly large differences between \Herwig and \Pythia for the quark-gluon separation as observed for \Herwig 2.7.1 in Refs.~\cite{Badger:2016bpw,Gras:2017jty}.)
This highlights the substantial improvement in the description of gluon jets in the latest version of \Herwig.

Finally, in \fig{classifier} we show the classifier separation at NLL$'$+NLO compared to \Pythia and \Herwig at parton and hadron level. This is similar to \fig{moneyplot}, but we do not impose a cut on thrust and therefore omit the NNLL$'$ result. The perturbative uncertainty $\de_{\rm pert}$ is shown, as well as the total uncertainty. Both \Pythia and \Herwig agree with our results within uncertainties. They differ from each other more than in \fig{moneyplot}, which is due to the relatively large differences in the gluon distribution at larger $\tau$. \Herwig predicts a lower classifier separation $\De$, because its gluon distribution is peaked at smaller values of $\tau$ and thus closer to the quark distribution. As in \fig{moneyplot}, this is most pronounced for the \Herwig dipole shower, which has the gluon distribution with the lowest peak and as a result gives the lowest $\De$.

Finally, it is worth noting that the resummation and hadronization uncertainties on the classifier separation are of similar size. Thus at higher orders the hadronization uncertainty currently becomes the limiting factor, as can be seen in the NNLL$'$ results in \fig{moneyplot}. This is of course also due to our rather generous variations for the hadronization parameter $\Omega_i$.
This situation can be improved by using a more refined treatment than carried out here, including renormalon subtractions and performing a fit to LEP data as done in Ref.~\cite{Abbate:2010xh}, which yields a much more precise determination of $\Omega_q$. However, one would then also have to perform a more careful treatment of the full shape function in the nonperturbative peak region of the quark distribution, for example using the methods of Refs.~\cite{Ligeti:2008ac, Bernlochner:2013gla}.

%%%%%%%%%%%%%%%%%%%%%%%%%%%%%%%%%%%%%%%%%%%%%%%%%%%%%%%%%%%%%%%%%%%%%%%%%%%%%%%%
\section{Conclusions}
\label{sec:conc}
%%%%%%%%%%%%%%%%%%%%%%%%%%%%%%%%%%%%%%%%%%%%%%%%%%%%%%%%%%%%%%%%%%%%%%%%%%%%%%%%

Large differences have been observed between parton showers in their prediction for our ability to discriminate quark jets from gluon jets. This inspired us to consider the thrust event shape, which can be calculated very precisely, obtaining a sample of quark jets from $Z \to q \bar q$ and gluon jets from $H \to gg$. We compared our analytic results up to NNLL$'$ to \Pythia and \Herwig, which represented the two opposite extremes in an earlier study~\cite{Badger:2016bpw,Gras:2017jty}. Our results are consistent with both \Pythia and \Herwig, though closest to \Pythia. This is due to the improved description of gluon jets in the most recent \Herwig release, while the previous \Herwig 7.0.4 showed substantial differences in the gluon distribution. Resummed predictions, like those obtained here, can thus serve as an important standard candle for parton showers. The perturbative uncertainties can be reduced further by going to higher orders. At NNLL$'$ the uncertainty from nonperturbative effects currently constitutes the limiting factor in the resummed results, which can be improved in the future with a more refined treatment of nonperturbative corrections.

%%%%%%%%%%%%%%%%%%%%%%%%%%%%%%%%%%%%%%%%%%%%%
\vspace{-1ex}
\begin{acknowledgments}
\vspace{-1ex}
We thank A.~Papaefstathiou, P.~Pietrulewicz and P.~Richardson for discussions.
F.T.~is supported by the DFG Emmy-Noether Grant No.~TA 867/1-1.
W.W.~is supported by the ERC grant ERC-STG-2015-677323 and the D-ITP consortium, a program of the Netherlands Organization for Scientific Research (NWO) that is funded by the Dutch Ministry of Education, Culture and Science (OCW). 
J.M.~thanks DESY for hospitality during the initial phase of this project.
\end{acknowledgments}

%%%%%%%%%%%%%%%%%%%%%%%%%%%%%%%%%%%%%%%%%%%%%

\appendix

%%%%%%%%%%%%%%%%%%%%%%%%%%%%%%%%%%%%%%%%%%%%%%%%%%%%%%%%%%%%%%%%%%%%%%%%%%%%%%%%
\section{Anomalous Dimensions}
%%%%%%%%%%%%%%%%%%%%%%%%%%%%%%%%%%%%%%%%%%%%%%%%%%%%%%%%%%%%%%%%%%%%%%%%%%%%%%%%

Expanding the beta function and anomalous dimensions in powers of $\alpha_s$,
%%%
\begin{align}
\beta(\alpha_s) &=
- 2 \alpha_s \sum_{n=0}^\infty \beta_n\Bigl(\frac{\alpha_s}{4\pi}\Bigr)^{n+1}
\,, \nn \\
\Gamma^i_\cusp(\alpha_s) &= \sum_{n=0}^\infty \Gamma^i_n \Bigl(\frac{\alpha_s}{4\pi}\Bigr)^{n+1}
\,, \nn \\
\gamma^i(\alpha_s) &= \sum_{n=0}^\infty \gamma^i_{n} \Bigl(\frac{\alpha_s}{4\pi}\Bigr)^{n+1}
\,,\end{align}
%%%
the coefficients are given by
%%%
\begin{widetext}
\begin{align}
\beta_0 &= \frac{11}{3}\,C_A -\frac{4}{3}\,T_F\,n_f
\,,\nn\\
\beta_1 &= \frac{34}{3}\,C_A^2  - \Bigl(\frac{20}{3}\,C_A\, + 4 C_F\Bigr)\, T_F\,n_f
\,, \nn\\
\beta_2 &=
\frac{2857}{54}\,C_A^3 + \Bigl(C_F^2 - \frac{205}{18}\,C_F C_A
 - \frac{1415}{54}\,C_A^2 \Bigr)\, 2T_F\,n_f
 + \Bigl(\frac{11}{9}\, C_F + \frac{79}{54}\, C_A \Bigr)\, 4T_F^2\,n_f^2
\,,\nn\\
\Gamma^q_0 &= 4C_F
\,,\nn\\
\Gamma^q_1 &= 4C_F \Bigl[\Bigl( \frac{67}{9} -\frac{\pi^2}{3} \Bigr)\,C_A  -
   \frac{20}{9}\,T_F\, n_f \Bigr]
\,,\nn\\
\Gamma^q_2 &= 4C_F \Bigl[
\Bigl(\frac{245}{6} \!-\! \frac{134 \pi^2}{27} \!+\! \frac{11 \pi ^4}{45}
  \!+\! \frac{22 \zeta_3}{3}\Bigr)C_A^2
  - \Bigl(\frac{418}{27} \!-\! \frac{40 \pi^2}{27}  \!+\! \frac{56 \zeta_3}{3} \Bigr)C_A\, T_F\,n_f
  - \Bigl(\frac{55}{3} \!-\! 16 \zeta_3 \Bigr) C_F\, T_F\,n_f
  - \frac{16}{27}\,T_F^2\, n_f^2 \Bigr]
\,,\nn \\
\gamma^q_{C\,0} &= -3 C_F
\,,\nn\\
\gamma^q_{C\,1}
&= - C_F \Bigl[
  \Bigl(\frac{41}{9} - 26 \zeta_3\Bigr) C_A
+ \Bigl(\frac32 - 2 \pi^2 + 24 \zeta_3\Bigr) C_F + \Bigl(\frac{65}{18} + \frac{\pi^2}{2} \Bigr) \beta_0 \Bigr]
\,, \nn \\
\gamma_{S\,0}^q &= 0
\,,\nn\\
\gamma_{S\,1}^q
&= C_F \Bigl[
  \Bigl(-\frac{128}{9} + 56 \zeta_3\Bigr) C_A
+ \Bigl(-\frac{112}{9} + \frac{2\pi^2}{3} \Bigr) \beta_0 \Bigr]
\,,\nn\\ 
\gamma_{C\,0}^g &= - \beta_0
\,,\nn\\
\gamma_{C\,1}^g
&= \Bigl(-\frac{59}{9} + 2\zeta_3\Bigr)C_A^2 +
\Bigl(-\frac{19}{9}+\frac{\pi^2}{6} \Bigr) C_A\, \beta_0 - \beta_1
\,. \end{align}
%%%
The expressions for $\Ga^g$ and $\ga_S^g$ are omitted, as they can be obtained from Casimir scaling
%%%
\begin{align}
  \Ga_n^g =  \frac{C_A}{C_F}\,\Ga_n^q
\,, \quad
  \ga_{S\,n}^g =  \frac{C_A}{C_F}\,\ga_{S\,n}^q  
\,.\end{align}
%%%
The coefficients $\ga_J^i$ of the non-cusp anomalous dimension of the jet function follow from \eq{consistency}.

%%%%%%%%%%%%%%%%%%%%%%%%%%%%%%%%%%%%%%%%%%%%%%%%%%%%%%%%%%%%%%%%%%%%%%%%%%%%%%%%
\section{Fixed-Order Ingredients}
%%%%%%%%%%%%%%%%%%%%%%%%%%%%%%%%%%%%%%%%%%%%%%%%%%%%%%%%%%%%%%%%%%%%%%%%%%%%%%%%

The form of the Wilson coefficient, jet function and soft function is highly constrained by the anomalous dimensions in \eq{RGE},
%%%
\begin{align}
  C_q(Q,\mu) &= 1 + \frac{\al_s(\mu)}{4\pi} \biggl[- \frac{\Ga_0^q}{4} \ln^2 \Bigl(\frac{-q^2-\img 0}{\mu^2}\Bigr) - \ga_{C\,0}^q \ln \Bigl(\frac{-q^2-\img 0}{\mu^2}\Bigr) + c_0^q \biggr] + \Bigl(\frac{\al_s(\mu)}{4\pi}\Bigr)^2 \biggl[\frac{(\Ga_0^q)^2}{32} \ln^4 \Bigl(\frac{-q^2-\img 0}{\mu^2}\Bigr) 
 \nn\\ & \quad\hspace{6ex}
 + \frac{\Ga_0^q(3\ga_{C\,0}^q + \bt_0)}{12} \ln^3 \Bigl(\frac{-q^2-\img 0}{\mu^2}\Bigr) + \frac{2(\ga_{C\,0}^q)^2 + 2 \bt_0 \ga_{C\,0}^q - \Ga_1^q - \Ga_0^q c_0^q}{4} \ln^2 \Bigl(\frac{-q^2-\img 0}{\mu^2}\Bigr)
 \nn\\ & \quad\hspace{6ex}
 -(\ga_{C\,1}^q + \ga_{C\,0}^q c_0^q + \bt_0 c_0^q) \ln \Bigl(\frac{-q^2-\img 0}{\mu^2}\Bigr) + c_1^q \biggr]
  \,,\nn \\ 
  C_g(Q,\mu) &= \al_s\biggl\{1 + \frac{\alpha_s(\mu)}{4\pi} \biggl[- \frac{\Ga_0^g}{4} \ln^2 \Bigl(\frac{-q^2-\img 0}{\mu^2}\Bigr) - (\ga_{C\,0}^g + \bt_0) \ln \Bigl(\frac{-q^2-\img 0}{\mu^2}\Bigr) + c_0^g \biggr]  + \Bigl(\frac{\al_s(\mu)}{4\pi}\Bigr)^2 \biggl[\frac{(\Ga_0^g)^2}{32} \ln^4 \Bigl(\frac{-q^2-\img 0}{\mu^2}\Bigr) 
 \nn\\ & \quad\hspace{6ex}
 + \frac{\Ga_0^g(3\ga_{C\,0}^g + 4\bt_0)}{12} \ln^3 \Bigl(\frac{-q^2-\img 0}{\mu^2}\Bigr) + \frac{2(\ga_{C\,0}^g)^2 + 6\bt_0 \ga_{C\,0}^g + 4\bt_0^2 - \Ga_1^g - \Ga_0^g c_0^g}{4} \ln^2 \Bigl(\frac{-q^2-\img 0}{\mu^2}\Bigr)
 \nn\\ & \quad\hspace{6ex}
 - (\ga_{C\,1}^g + \ga_{C\,0}^g c_0^g + 2\bt_0 c_0^g + \bt_1) \ln \Bigl(\frac{-q^2-\img 0}{\mu^2}\Bigr) + c_1^g \biggr\}
 \,, \nn \\ 
 J_i(s,\mu) &= \de(s) + \frac{\al_s(\mu)}{4\pi} \biggl[ \frac{\Ga_0^i}{\mu^2} \Big(\frac{\ln(s/\mu^2)}{(s/\mu^2)}\Big)_+ - \frac{\ga_{J\,0}^i}{2 \mu^2} \frac{1}{(s/\mu^2)}_+ + j_0^i\, \de(s) \biggr] + \Bigl(\frac{\al_s(\mu)}{4\pi}\Bigr)^2 \biggl[\frac{(\Ga_0^i)^2}{2\mu^2} \Bigl(\frac{\ln^3(s/\mu^2)}{(s/\mu^2)}\Bigr)_+
 \nn\\ & \quad\hspace{6ex}
  -\Ga_0^i \frac{2\bt_0 + 3\ga_{J\,0}^i}{4\mu^2} \Bigl(\frac{\ln^2 (s/\mu^2)}{(s/\mu^2)}\Bigr)_+ + \Bigl(j_0^i \Ga_0^i + \frac{\bt_0 \ga_{J\,0}^i}{2} + \frac{(\ga_{J\,0}^i)^2}{4} - \frac{\pi^2 (\Ga_0^i)^2}{6} + \Ga_1^i \Bigr) \frac{1}{\mu^2} \Bigl(\frac{\ln(s/\mu^2)}{(s/\mu^2)}\Bigr)_+
 \nn\\ & \quad\hspace{6ex}
   + \Bigl(-j_0^i\Bigl(\bt_0 + \frac{\ga_{J\,0}^i}{2}\Bigr) - \frac{\ga_{J\,1}^i}{2} + \frac{\pi^2 \ga_{J\,0}^i \Ga_0^i}{12} + \zeta_3 (\Ga_0^i)^2 \Bigr) \frac{1}{\mu^2} \frac{1}{(s/\mu^2)_+} + j_1^i \de(s) \biggr]
 \,,\nn \\
 S_i(k,\mu) &= \delta(k) + \frac{\alpha_s(\mu)}{4\pi} \biggl[
-\frac{4\Ga_0^i}{\mu} \Bigl(\frac{\ln(k/\mu)}{(k/\mu)}\Bigr)_+ -
\frac{\ga_{S\,0}^i}{\mu} \Bigl(\frac{1}{(k/\mu)}\Bigr)_+ + s_0^i\, \delta(k) \biggr] + \Bigl(\frac{\al_s(\mu)}{4\pi}\Bigr)^2 \biggl[\frac{8(\Ga_0^i)^2}{\mu} \Bigl(\frac{\ln^3(k/\mu)}{(k/\mu)}\Bigr)_+
 \nn\\ & \quad\hspace{6ex}
 + \Ga_0^i \frac{4\bt_0 + 6\ga_{S\,0}^i}{\mu} \Bigl(\frac{\ln^2(k/\mu)}{(k/\mu)}\Bigr)_+ + \Bigl(- 4 s_0^i \Ga_0^i + 2\bt_0 \ga_{S\,0}^i + (\ga_{S\,0}^i)^2 - \frac{8\pi^2 (\Ga_0^i)^2}{3} - 4\Ga_1^i\Bigr) \frac{1}{\mu} \Bigl(\frac{\ln(k/\mu)}{(k/\mu)}\Bigr)_+
  \nn\\ & \quad\hspace{6ex}
  + \Bigl(-s_0^i\Bigl(2\bt_0 + \ga_{S\,0}^i\Bigr) - \ga_{S\,1}^i - \frac{2\pi^2 \ga_{S\,0}^i \Ga_0^i}{3} + 16\zeta_3 (\Ga_0^i)^2 \Bigr) \frac{1}{\mu} \frac{1}{(k/\mu)_+} + s_1^i \de(k) \biggr]
\,.\end{align}
%%%
The difference between the expressions for $C_q$ and $C_g$ is due to the additional prefactor of $\al_s$ in the latter. The remaining constants are given by
%%%
\begin{align}
  c_0^q &= \Bigl(- 8 + \frac{\pi^2}{6}\Bigr) C_F
  \,, \nn \\
  c_1^q &= \Bigl(\frac{255}{8} + \frac{7\pi^2}{2} - 30\zeta_3 - \frac{83\pi^4}{360}\Bigr) C_F^2 + \Bigl(-\frac{51157}{648} - \frac{337\pi^2}{108} + \frac{313\zeta_3}{9} + \frac{11\pi^4}{45}\Bigr) C_F C_A + \Bigl(\frac{4085}{162} + \frac{23\pi^2}{27} +\frac{4\zeta_3}{9}\Bigr) C_F T_F n_f
  \,, \nn \\
  c_0^g &= \Bigl(5 + \frac{\pi^2}{6}\Bigr) C_A - 3 C_F
  \,, \nn \\
  c_1^g &= (7C_A^2 + 11C_A C_F - 6C_F \bt_0)\ln\Bigl(\frac{-q^2-i0}{m_t^2}\Bigr) + \Bigl(-\frac{419}{27} + \frac{7\pi^2}{6} + \frac{\pi^4}{72} - 44\zeta_3\Bigr)C_A^2 + \Bigl(-\frac{217}{2} - \frac{\pi^2}{2} + 44\zeta_3\Bigr)C_A C_F
     \nn\\ & \quad\hspace{6ex}
     + \Bigl(\frac{2255}{108} + \frac{5\pi^2}{12} + \frac{23\zeta_3}{3}\Bigr)C_A \bt_0 - \frac{5}{6}C_A T_F + \frac{27}{2} C_F^2 + \Bigl(\frac{41}{2} - 12\zeta_3\Bigr)C_F \bt_0 -\frac{4}{3}C_F T_F + \mathcal{O}\Bigl(\frac{q^2}{4m_t^2}\Bigr)
  \,, \nn \\
  j_0^q &= (7-\pi^2) C_F
  \,, \nn \\
  j_1^q &= \Bigl(\frac{205}{8} - \frac{67\pi^2}{6} + \frac{14\pi^4}{15} - 18\zeta^3 \Bigr) C_F^2 + \Bigl(\frac{53129}{648} - \frac{208\pi^2}{27} - \frac{17\pi^4}{180} - \frac{206\zeta_3}{9} \Bigr) C_F C_A + \Bigl(\frac{4057}{162} + \frac{68\pi^2}{27} + \frac{16\zeta_3}{9}\Bigr) C_F T_F n_f
  \,, \nn \\
  j_0^g &= \Bigl(\frac{4}{3} - \pi^2 \Bigr)C_A + \frac{5}{3} \bt_0
  \,, \nn \\
  j_1^g &= \Bigl(\frac{4255}{108} - \frac{26\pi^2}{9} + \frac{151\pi^4}{180} - 72\zeta^3 \Bigr) C_A^2 - \Bigl(\frac{115}{108} + \frac{65\pi^2}{18} - \frac{56\zeta_3}{3} \Bigr) C_A \bt_0 - \Bigl(\frac{25}{9}-\frac{\pi^2}{3}\Bigr) \bt_0^2 + \Bigl(\frac{55}{12} - 4\zeta_3\Bigr) \bt_0^2
  \,, \nn \\
  s_0^q &= \frac{\pi^2}{3} C_F  
  \,, \nn \\
  s_1^q &= \frac{-3\pi^4}{10} C_F^2 + \Bigl(\frac{-640}{27} + \frac{4\pi^2}{3} + \frac{22\pi^4}{45}\Bigr) C_F C_A + \Bigl(\frac{-20}{27} - \frac{37\pi^2}{18} + \frac{58\zeta_3}{3}\Bigr) C_F \bt_0
\,.\end{align}
%%%
The coefficients $s_n^g$ for the gluon soft function can directly be obtained from $s_n^q$ by replacing $C_F \to C_A$.

\end{widetext}

\bibliography{q_vs_g}

\begin{thebibliography}{53}%
\makeatletter
\providecommand \@ifxundefined [1]{%
 \@ifx{#1\undefined}
}%
\providecommand \@ifnum [1]{%
 \ifnum #1\expandafter \@firstoftwo
 \else \expandafter \@secondoftwo
 \fi
}%
\providecommand \@ifx [1]{%
 \ifx #1\expandafter \@firstoftwo
 \else \expandafter \@secondoftwo
 \fi
}%
\providecommand \natexlab [1]{#1}%
\providecommand \enquote  [1]{``#1''}%
\providecommand \bibnamefont  [1]{#1}%
\providecommand \bibfnamefont [1]{#1}%
\providecommand \citenamefont [1]{#1}%
\providecommand \href@noop [0]{\@secondoftwo}%
\providecommand \href [0]{\begingroup \@sanitize@url \@href}%
\providecommand \@href[1]{\@@startlink{#1}\@@href}%
\providecommand \@@href[1]{\endgroup#1\@@endlink}%
\providecommand \@sanitize@url [0]{\catcode `\\12\catcode `\$12\catcode
  `\&12\catcode `\#12\catcode `\^12\catcode `\_12\catcode `\%12\relax}%
\providecommand \@@startlink[1]{}%
\providecommand \@@endlink[0]{}%
\providecommand \url  [0]{\begingroup\@sanitize@url \@url }%
\providecommand \@url [1]{\endgroup\@href {#1}{\urlprefix }}%
\providecommand \urlprefix  [0]{URL }%
\providecommand \Eprint [0]{\href }%
\providecommand \doibase [0]{http://dx.doi.org/}%
\providecommand \selectlanguage [0]{\@gobble}%
\providecommand \bibinfo  [0]{\@secondoftwo}%
\providecommand \bibfield  [0]{\@secondoftwo}%
\providecommand \translation [1]{[#1]}%
\providecommand \BibitemOpen [0]{}%
\providecommand \bibitemStop [0]{}%
\providecommand \bibitemNoStop [0]{.\EOS\space}%
\providecommand \EOS [0]{\spacefactor3000\relax}%
\providecommand \BibitemShut  [1]{\csname bibitem#1\endcsname}%
\let\auto@bib@innerbib\@empty
%</preamble>
\bibitem [{\citenamefont {Abdesselam}\ \emph {et~al.}(2011)\citenamefont
  {Abdesselam} \emph {et~al.}}]{Abdesselam:2010pt}%
  \BibitemOpen
  \bibfield  {author} {\bibinfo {author} {\bibfnamefont {A.}~\bibnamefont
  {Abdesselam}} \emph {et~al.},\ }\href {\doibase
  10.1140/epjc/s10052-011-1661-y} {\bibfield  {journal} {\bibinfo  {journal}
  {Eur. Phys. J.}\ }\textbf {\bibinfo {volume} {C71}},\ \bibinfo {pages} {1661}
  (\bibinfo {year} {2011})},\ \Eprint {http://arxiv.org/abs/1012.5412}
  {arXiv:1012.5412 [hep-ph]} \BibitemShut {NoStop}%
%%CITATION = ARXIV:1012.5412;%%
\bibitem [{\citenamefont {Altheimer}\ \emph {et~al.}(2012)\citenamefont
  {Altheimer} \emph {et~al.}}]{Altheimer:2012mn}%
  \BibitemOpen
  \bibfield  {author} {\bibinfo {author} {\bibfnamefont {A.}~\bibnamefont
  {Altheimer}} \emph {et~al.},\ }\href {\doibase 10.1088/0954-3899/39/6/063001}
  {\bibfield  {journal} {\bibinfo  {journal} {J. Phys.}\ }\textbf {\bibinfo
  {volume} {G39}},\ \bibinfo {pages} {063001} (\bibinfo {year} {2012})},\
  \Eprint {http://arxiv.org/abs/1201.0008} {arXiv:1201.0008 [hep-ph]}
  \BibitemShut {NoStop}%
%%CITATION = ARXIV:1201.0008;%%
\bibitem [{\citenamefont {Altheimer}\ \emph {et~al.}(2014)\citenamefont
  {Altheimer} \emph {et~al.}}]{Altheimer:2013yza}%
  \BibitemOpen
  \bibfield  {author} {\bibinfo {author} {\bibfnamefont {A.}~\bibnamefont
  {Altheimer}} \emph {et~al.},\ }\href {\doibase
  10.1140/epjc/s10052-014-2792-8} {\bibfield  {journal} {\bibinfo  {journal}
  {Eur. Phys. J.}\ }\textbf {\bibinfo {volume} {C74}},\ \bibinfo {pages} {2792}
  (\bibinfo {year} {2014})},\ \Eprint {http://arxiv.org/abs/1311.2708}
  {arXiv:1311.2708 [hep-ex]} \BibitemShut {NoStop}%
%%CITATION = ARXIV:1311.2708;%%
\bibitem [{\citenamefont {Adams}\ \emph {et~al.}(2015)\citenamefont {Adams}
  \emph {et~al.}}]{Adams:2015hiv}%
  \BibitemOpen
  \bibfield  {author} {\bibinfo {author} {\bibfnamefont {D.}~\bibnamefont
  {Adams}} \emph {et~al.},\ }\href {\doibase 10.1140/epjc/s10052-015-3587-2}
  {\bibfield  {journal} {\bibinfo  {journal} {Eur. Phys. J.}\ }\textbf
  {\bibinfo {volume} {C75}},\ \bibinfo {pages} {409} (\bibinfo {year}
  {2015})},\ \Eprint {http://arxiv.org/abs/1504.00679} {arXiv:1504.00679
  [hep-ph]} \BibitemShut {NoStop}%
%%CITATION = ARXIV:1504.00679;%%
\bibitem [{\citenamefont {Gallicchio}\ and\ \citenamefont
  {Schwartz}(2011)}]{Gallicchio:2011xq}%
  \BibitemOpen
  \bibfield  {author} {\bibinfo {author} {\bibfnamefont {J.}~\bibnamefont
  {Gallicchio}}\ and\ \bibinfo {author} {\bibfnamefont {M.~D.}\ \bibnamefont
  {Schwartz}},\ }\href {\doibase 10.1103/PhysRevLett.107.172001} {\bibfield
  {journal} {\bibinfo  {journal} {Phys. Rev. Lett.}\ }\textbf {\bibinfo
  {volume} {107}},\ \bibinfo {pages} {172001} (\bibinfo {year} {2011})},\
  \Eprint {http://arxiv.org/abs/1106.3076} {arXiv:1106.3076 [hep-ph]}
  \BibitemShut {NoStop}%
%%CITATION = ARXIV:1106.3076;%%
\bibitem [{\citenamefont {Gallicchio}\ and\ \citenamefont
  {Schwartz}(2013)}]{Gallicchio:2012ez}%
  \BibitemOpen
  \bibfield  {author} {\bibinfo {author} {\bibfnamefont {J.}~\bibnamefont
  {Gallicchio}}\ and\ \bibinfo {author} {\bibfnamefont {M.~D.}\ \bibnamefont
  {Schwartz}},\ }\href {\doibase 10.1007/JHEP04(2013)090} {\bibfield  {journal}
  {\bibinfo  {journal} {JHEP}\ }\textbf {\bibinfo {volume} {04}},\ \bibinfo
  {pages} {090} (\bibinfo {year} {2013})},\ \Eprint
  {http://arxiv.org/abs/1211.7038} {arXiv:1211.7038 [hep-ph]} \BibitemShut
  {NoStop}%
%%CITATION = ARXIV:1211.7038;%%
\bibitem [{\citenamefont {Larkoski}\ \emph {et~al.}(2013)\citenamefont
  {Larkoski}, \citenamefont {Salam},\ and\ \citenamefont
  {Thaler}}]{Larkoski:2013eya}%
  \BibitemOpen
  \bibfield  {author} {\bibinfo {author} {\bibfnamefont {A.~J.}\ \bibnamefont
  {Larkoski}}, \bibinfo {author} {\bibfnamefont {G.~P.}\ \bibnamefont {Salam}},
  \ and\ \bibinfo {author} {\bibfnamefont {J.}~\bibnamefont {Thaler}},\ }\href
  {\doibase 10.1007/JHEP06(2013)108} {\bibfield  {journal} {\bibinfo  {journal}
  {JHEP}\ }\textbf {\bibinfo {volume} {06}},\ \bibinfo {pages} {108} (\bibinfo
  {year} {2013})},\ \Eprint {http://arxiv.org/abs/1305.0007} {arXiv:1305.0007
  [hep-ph]} \BibitemShut {NoStop}%
%%CITATION = ARXIV:1305.0007;%%
\bibitem [{\citenamefont {Larkoski}\ \emph {et~al.}(2014)\citenamefont
  {Larkoski}, \citenamefont {Thaler},\ and\ \citenamefont
  {Waalewijn}}]{Larkoski:2014pca}%
  \BibitemOpen
  \bibfield  {author} {\bibinfo {author} {\bibfnamefont {A.~J.}\ \bibnamefont
  {Larkoski}}, \bibinfo {author} {\bibfnamefont {J.}~\bibnamefont {Thaler}}, \
  and\ \bibinfo {author} {\bibfnamefont {W.~J.}\ \bibnamefont {Waalewijn}},\
  }\href {\doibase 10.1007/JHEP11(2014)129} {\bibfield  {journal} {\bibinfo
  {journal} {JHEP}\ }\textbf {\bibinfo {volume} {11}},\ \bibinfo {pages} {129}
  (\bibinfo {year} {2014})},\ \Eprint {http://arxiv.org/abs/1408.3122}
  {arXiv:1408.3122 [hep-ph]} \BibitemShut {NoStop}%
%%CITATION = ARXIV:1408.3122;%%
\bibitem [{\citenamefont {Bhattacherjee}\ \emph {et~al.}(2015)\citenamefont
  {Bhattacherjee}, \citenamefont {Mukhopadhyay}, \citenamefont {Nojiri},
  \citenamefont {Sakaki},\ and\ \citenamefont
  {Webber}}]{Bhattacherjee:2015psa}%
  \BibitemOpen
  \bibfield  {author} {\bibinfo {author} {\bibfnamefont {B.}~\bibnamefont
  {Bhattacherjee}}, \bibinfo {author} {\bibfnamefont {S.}~\bibnamefont
  {Mukhopadhyay}}, \bibinfo {author} {\bibfnamefont {M.~M.}\ \bibnamefont
  {Nojiri}}, \bibinfo {author} {\bibfnamefont {Y.}~\bibnamefont {Sakaki}}, \
  and\ \bibinfo {author} {\bibfnamefont {B.~R.}\ \bibnamefont {Webber}},\
  }\href {\doibase 10.1007/JHEP04(2015)131} {\bibfield  {journal} {\bibinfo
  {journal} {JHEP}\ }\textbf {\bibinfo {volume} {04}},\ \bibinfo {pages} {131}
  (\bibinfo {year} {2015})},\ \Eprint {http://arxiv.org/abs/1501.04794}
  {arXiv:1501.04794 [hep-ph]} \BibitemShut {NoStop}%
%%CITATION = ARXIV:1501.04794;%%
\bibitem [{\citenamefont {Andersen}\ \emph {et~al.}(2016)\citenamefont
  {Andersen} \emph {et~al.}}]{Badger:2016bpw}%
  \BibitemOpen
  \bibfield  {author} {\bibinfo {author} {\bibfnamefont {J.~R.}\ \bibnamefont
  {Andersen}} \emph {et~al.},\ }in\ \href
  {http://lss.fnal.gov/archive/2016/conf/fermilab-conf-16-175-ppd-t.pdf} {\emph
  {\bibinfo {booktitle} {{9th Les Houches Workshop on Physics at TeV Colliders
  (PhysTeV 2015) Les Houches, France, June 1-19, 2015}}}}\ (\bibinfo {year}
  {2016})\ \Eprint {http://arxiv.org/abs/1605.04692} {arXiv:1605.04692
  [hep-ph]} \BibitemShut {NoStop}%
%%CITATION = ARXIV:1605.04692;%%
\bibitem [{\citenamefont {Ferreira~de Lima}\ \emph {et~al.}(2017)\citenamefont
  {Ferreira~de Lima}, \citenamefont {Petrov}, \citenamefont {Soper},\ and\
  \citenamefont {Spannowsky}}]{FerreiradeLima:2016gcz}%
  \BibitemOpen
  \bibfield  {author} {\bibinfo {author} {\bibfnamefont {D.}~\bibnamefont
  {Ferreira~de Lima}}, \bibinfo {author} {\bibfnamefont {P.}~\bibnamefont
  {Petrov}}, \bibinfo {author} {\bibfnamefont {D.}~\bibnamefont {Soper}}, \
  and\ \bibinfo {author} {\bibfnamefont {M.}~\bibnamefont {Spannowsky}},\
  }\href {\doibase 10.1103/PhysRevD.95.034001} {\bibfield  {journal} {\bibinfo
  {journal} {Phys. Rev.}\ }\textbf {\bibinfo {volume} {D95}},\ \bibinfo {pages}
  {034001} (\bibinfo {year} {2017})},\ \Eprint
  {http://arxiv.org/abs/1607.06031} {arXiv:1607.06031 [hep-ph]} \BibitemShut
  {NoStop}%
%%CITATION = ARXIV:1607.06031;%%
\bibitem [{\citenamefont {Komiske}\ \emph {et~al.}(2017)\citenamefont
  {Komiske}, \citenamefont {Metodiev},\ and\ \citenamefont
  {Schwartz}}]{Komiske:2016rsd}%
  \BibitemOpen
  \bibfield  {author} {\bibinfo {author} {\bibfnamefont {P.~T.}\ \bibnamefont
  {Komiske}}, \bibinfo {author} {\bibfnamefont {E.~M.}\ \bibnamefont
  {Metodiev}}, \ and\ \bibinfo {author} {\bibfnamefont {M.~D.}\ \bibnamefont
  {Schwartz}},\ }\href {\doibase 10.1007/JHEP01(2017)110} {\bibfield  {journal}
  {\bibinfo  {journal} {JHEP}\ }\textbf {\bibinfo {volume} {01}},\ \bibinfo
  {pages} {110} (\bibinfo {year} {2017})},\ \Eprint
  {http://arxiv.org/abs/1612.01551} {arXiv:1612.01551 [hep-ph]} \BibitemShut
  {NoStop}%
%%CITATION = ARXIV:1612.01551;%%
\bibitem [{\citenamefont {Davighi}\ and\ \citenamefont
  {Harris}(2017)}]{Davighi:2017hok}%
  \BibitemOpen
  \bibfield  {author} {\bibinfo {author} {\bibfnamefont {J.}~\bibnamefont
  {Davighi}}\ and\ \bibinfo {author} {\bibfnamefont {P.}~\bibnamefont
  {Harris}},\ }\href@noop {} {\  (\bibinfo {year} {2017})},\ \Eprint
  {http://arxiv.org/abs/1703.00914} {arXiv:1703.00914 [hep-ph]} \BibitemShut
  {NoStop}%
%%CITATION = ARXIV:1703.00914;%%
\bibitem [{\citenamefont {Gras}\ \emph {et~al.}(2017)\citenamefont {Gras},
  \citenamefont {Hoeche}, \citenamefont {Kar}, \citenamefont {Larkoski},
  \citenamefont {Lšnnblad}, \citenamefont {PlŠtzer}, \citenamefont {Si—dmok},
  \citenamefont {Skands}, \citenamefont {Soyez},\ and\ \citenamefont
  {Thaler}}]{Gras:2017jty}%
  \BibitemOpen
  \bibfield  {author} {\bibinfo {author} {\bibfnamefont {P.}~\bibnamefont
  {Gras}}, \bibinfo {author} {\bibfnamefont {S.}~\bibnamefont {Hoeche}},
  \bibinfo {author} {\bibfnamefont {D.}~\bibnamefont {Kar}}, \bibinfo {author}
  {\bibfnamefont {A.}~\bibnamefont {Larkoski}}, \bibinfo {author}
  {\bibfnamefont {L.}~\bibnamefont {Lšnnblad}}, \bibinfo {author}
  {\bibfnamefont {S.}~\bibnamefont {PlŠtzer}}, \bibinfo {author} {\bibfnamefont
  {A.}~\bibnamefont {Si—dmok}}, \bibinfo {author} {\bibfnamefont
  {P.}~\bibnamefont {Skands}}, \bibinfo {author} {\bibfnamefont
  {G.}~\bibnamefont {Soyez}}, \ and\ \bibinfo {author} {\bibfnamefont
  {J.}~\bibnamefont {Thaler}},\ }\href@noop {} {\  (\bibinfo {year} {2017})},\
  \Eprint {http://arxiv.org/abs/1704.03878} {arXiv:1704.03878 [hep-ph]}
  \BibitemShut {NoStop}%
%%CITATION = ARXIV:1704.03878;%%
\bibitem [{\citenamefont {Sj{\"o}strand}\ \emph {et~al.}(2015)\citenamefont
  {Sj{\"o}strand}, \citenamefont {Ask}, \citenamefont {Christiansen},
  \citenamefont {Corke}, \citenamefont {Desai}, \citenamefont {Ilten},
  \citenamefont {Mrenna}, \citenamefont {Prestel}, \citenamefont {Rasmussen},\
  and\ \citenamefont {Skands}}]{Sjostrand:2014zea}%
  \BibitemOpen
  \bibfield  {author} {\bibinfo {author} {\bibfnamefont {T.}~\bibnamefont
  {Sj{\"o}strand}}, \bibinfo {author} {\bibfnamefont {S.}~\bibnamefont {Ask}},
  \bibinfo {author} {\bibfnamefont {J.~R.}\ \bibnamefont {Christiansen}},
  \bibinfo {author} {\bibfnamefont {R.}~\bibnamefont {Corke}}, \bibinfo
  {author} {\bibfnamefont {N.}~\bibnamefont {Desai}}, \bibinfo {author}
  {\bibfnamefont {P.}~\bibnamefont {Ilten}}, \bibinfo {author} {\bibfnamefont
  {S.}~\bibnamefont {Mrenna}}, \bibinfo {author} {\bibfnamefont
  {S.}~\bibnamefont {Prestel}}, \bibinfo {author} {\bibfnamefont {C.~O.}\
  \bibnamefont {Rasmussen}}, \ and\ \bibinfo {author} {\bibfnamefont {P.~Z.}\
  \bibnamefont {Skands}},\ }\href {\doibase 10.1016/j.cpc.2015.01.024}
  {\bibfield  {journal} {\bibinfo  {journal} {Comput. Phys. Commun.}\ }\textbf
  {\bibinfo {volume} {191}},\ \bibinfo {pages} {159} (\bibinfo {year}
  {2015})},\ \Eprint {http://arxiv.org/abs/1410.3012} {arXiv:1410.3012
  [hep-ph]} \BibitemShut {NoStop}%
%%CITATION = ARXIV:1410.3012;%%
\bibitem [{\citenamefont {Bellm}\ \emph {et~al.}(2017)\citenamefont {Bellm}
  \emph {et~al.}}]{Bellm:2017bvx}%
  \BibitemOpen
  \bibfield  {author} {\bibinfo {author} {\bibfnamefont {J.}~\bibnamefont
  {Bellm}} \emph {et~al.},\ }\href@noop {} {\  (\bibinfo {year} {2017})},\
  \Eprint {http://arxiv.org/abs/1705.06919} {arXiv:1705.06919 [hep-ph]}
  \BibitemShut {NoStop}%
%%CITATION = ARXIV:1705.06919;%%
\bibitem [{\citenamefont {Becher}\ and\ \citenamefont
  {Schwartz}(2008)}]{Becher:2008cf}%
  \BibitemOpen
  \bibfield  {author} {\bibinfo {author} {\bibfnamefont {T.}~\bibnamefont
  {Becher}}\ and\ \bibinfo {author} {\bibfnamefont {M.~D.}\ \bibnamefont
  {Schwartz}},\ }\href {\doibase 10.1088/1126-6708/2008/07/034} {\bibfield
  {journal} {\bibinfo  {journal} {JHEP}\ }\textbf {\bibinfo {volume} {07}},\
  \bibinfo {pages} {034} (\bibinfo {year} {2008})},\ \Eprint
  {http://arxiv.org/abs/0803.0342} {arXiv:0803.0342 [hep-ph]} \BibitemShut
  {NoStop}%
%%CITATION = ARXIV:0803.0342;%%
\bibitem [{\citenamefont {Abbate}\ \emph {et~al.}(2011)\citenamefont {Abbate},
  \citenamefont {Fickinger}, \citenamefont {Hoang}, \citenamefont {Mateu},\
  and\ \citenamefont {Stewart}}]{Abbate:2010xh}%
  \BibitemOpen
  \bibfield  {author} {\bibinfo {author} {\bibfnamefont {R.}~\bibnamefont
  {Abbate}}, \bibinfo {author} {\bibfnamefont {M.}~\bibnamefont {Fickinger}},
  \bibinfo {author} {\bibfnamefont {A.~H.}\ \bibnamefont {Hoang}}, \bibinfo
  {author} {\bibfnamefont {V.}~\bibnamefont {Mateu}}, \ and\ \bibinfo {author}
  {\bibfnamefont {I.~W.}\ \bibnamefont {Stewart}},\ }\href {\doibase
  10.1103/PhysRevD.83.074021} {\bibfield  {journal} {\bibinfo  {journal} {Phys.
  Rev.}\ }\textbf {\bibinfo {volume} {D83}},\ \bibinfo {pages} {074021}
  (\bibinfo {year} {2011})},\ \Eprint {http://arxiv.org/abs/1006.3080}
  {arXiv:1006.3080 [hep-ph]} \BibitemShut {NoStop}%
%%CITATION = ARXIV:1006.3080;%%
\bibitem [{\citenamefont {Korchemsky}\ and\ \citenamefont
  {Sterman}(1999)}]{Korchemsky:1999kt}%
  \BibitemOpen
  \bibfield  {author} {\bibinfo {author} {\bibfnamefont {G.~P.}\ \bibnamefont
  {Korchemsky}}\ and\ \bibinfo {author} {\bibfnamefont {G.~F.}\ \bibnamefont
  {Sterman}},\ }\href {\doibase 10.1016/S0550-3213(99)00308-9} {\bibfield
  {journal} {\bibinfo  {journal} {Nucl. Phys.}\ }\textbf {\bibinfo {volume}
  {B555}},\ \bibinfo {pages} {335} (\bibinfo {year} {1999})},\ \Eprint
  {http://arxiv.org/abs/hep-ph/9902341} {arXiv:hep-ph/9902341} \BibitemShut
  {NoStop}%
%%CITATION = HEP-PH/9902341;%%
\bibitem [{\citenamefont {Korchemsky}\ and\ \citenamefont
  {Tafat}(2000)}]{Korchemsky:2000kp}%
  \BibitemOpen
  \bibfield  {author} {\bibinfo {author} {\bibfnamefont {G.~P.}\ \bibnamefont
  {Korchemsky}}\ and\ \bibinfo {author} {\bibfnamefont {S.}~\bibnamefont
  {Tafat}},\ }\href {\doibase 10.1088/1126-6708/2000/10/010} {\bibfield
  {journal} {\bibinfo  {journal} {JHEP}\ }\textbf {\bibinfo {volume} {10}},\
  \bibinfo {pages} {010} (\bibinfo {year} {2000})},\ \Eprint
  {http://arxiv.org/abs/hep-ph/0007005} {arXiv:hep-ph/0007005} \BibitemShut
  {NoStop}%
%%CITATION = HEP-PH/0007005;%%
\bibitem [{\citenamefont {Hoang}\ and\ \citenamefont
  {Stewart}(2008)}]{Hoang:2007vb}%
  \BibitemOpen
  \bibfield  {author} {\bibinfo {author} {\bibfnamefont {A.~H.}\ \bibnamefont
  {Hoang}}\ and\ \bibinfo {author} {\bibfnamefont {I.~W.}\ \bibnamefont
  {Stewart}},\ }\href {\doibase 10.1016/j.physletb.2008.01.040} {\bibfield
  {journal} {\bibinfo  {journal} {Phys. Lett.}\ }\textbf {\bibinfo {volume}
  {B660}},\ \bibinfo {pages} {483} (\bibinfo {year} {2008})},\ \Eprint
  {http://arxiv.org/abs/0709.3519} {arXiv:0709.3519 [hep-ph]} \BibitemShut
  {NoStop}%
%%CITATION = ARXIV:0709.3519;%%
\bibitem [{\citenamefont {Ligeti}\ \emph {et~al.}(2008)\citenamefont {Ligeti},
  \citenamefont {Stewart},\ and\ \citenamefont {Tackmann}}]{Ligeti:2008ac}%
  \BibitemOpen
  \bibfield  {author} {\bibinfo {author} {\bibfnamefont {Z.}~\bibnamefont
  {Ligeti}}, \bibinfo {author} {\bibfnamefont {I.~W.}\ \bibnamefont {Stewart}},
  \ and\ \bibinfo {author} {\bibfnamefont {F.~J.}\ \bibnamefont {Tackmann}},\
  }\href {\doibase 10.1103/PhysRevD.78.114014} {\bibfield  {journal} {\bibinfo
  {journal} {Phys. Rev.}\ }\textbf {\bibinfo {volume} {D78}},\ \bibinfo {pages}
  {114014} (\bibinfo {year} {2008})},\ \Eprint {http://arxiv.org/abs/0807.1926}
  {arXiv:0807.1926 [hep-ph]} \BibitemShut {NoStop}%
%%CITATION = ARXIV:0807.1926;%%
\bibitem [{\citenamefont {Richardson}()}]{Richardson}%
  \BibitemOpen
  \bibfield  {author} {\bibinfo {author} {\bibfnamefont {P.}~\bibnamefont
  {Richardson}},\ }\href@noop {} {\bibinfo  {journal} {private communication}\
  }\BibitemShut {NoStop}%
\bibitem [{\citenamefont {Catani}\ \emph {et~al.}(1993)\citenamefont {Catani},
  \citenamefont {Trentadue}, \citenamefont {Turnock},\ and\ \citenamefont
  {Webber}}]{Catani:1992ua}%
  \BibitemOpen
\bibfield  {journal} {  }\bibfield  {author} {\bibinfo {author} {\bibfnamefont
  {S.}~\bibnamefont {Catani}}, \bibinfo {author} {\bibfnamefont
  {L.}~\bibnamefont {Trentadue}}, \bibinfo {author} {\bibfnamefont
  {G.}~\bibnamefont {Turnock}}, \ and\ \bibinfo {author} {\bibfnamefont
  {B.~R.}\ \bibnamefont {Webber}},\ }\href {\doibase
  10.1016/0550-3213(93)90271-P} {\bibfield  {journal} {\bibinfo  {journal}
  {Nucl. Phys.}\ }\textbf {\bibinfo {volume} {B407}},\ \bibinfo {pages} {3}
  (\bibinfo {year} {1993})}\BibitemShut {NoStop}%
%%CITATION = NUPHA,B407,3;%%
\bibitem [{\citenamefont {Fleming}\ \emph
  {et~al.}(2008{\natexlab{a}})\citenamefont {Fleming}, \citenamefont {Hoang},
  \citenamefont {Mantry},\ and\ \citenamefont {Stewart}}]{Fleming:2007qr}%
  \BibitemOpen
  \bibfield  {author} {\bibinfo {author} {\bibfnamefont {S.}~\bibnamefont
  {Fleming}}, \bibinfo {author} {\bibfnamefont {A.~H.}\ \bibnamefont {Hoang}},
  \bibinfo {author} {\bibfnamefont {S.}~\bibnamefont {Mantry}}, \ and\ \bibinfo
  {author} {\bibfnamefont {I.~W.}\ \bibnamefont {Stewart}},\ }\href {\doibase
  10.1103/PhysRevD.77.074010} {\bibfield  {journal} {\bibinfo  {journal} {Phys.
  Rev.}\ }\textbf {\bibinfo {volume} {D77}},\ \bibinfo {pages} {074010}
  (\bibinfo {year} {2008}{\natexlab{a}})},\ \Eprint
  {http://arxiv.org/abs/hep-ph/0703207} {arXiv:hep-ph/0703207} \BibitemShut
  {NoStop}%
%%CITATION = HEP-PH/0703207;%%
\bibitem [{\citenamefont {Schwartz}(2008)}]{Schwartz:2007ib}%
  \BibitemOpen
  \bibfield  {author} {\bibinfo {author} {\bibfnamefont {M.~D.}\ \bibnamefont
  {Schwartz}},\ }\href {\doibase 10.1103/PhysRevD.77.014026} {\bibfield
  {journal} {\bibinfo  {journal} {Phys. Rev.}\ }\textbf {\bibinfo {volume}
  {D77}},\ \bibinfo {pages} {014026} (\bibinfo {year} {2008})},\ \Eprint
  {http://arxiv.org/abs/0709.2709} {arXiv:0709.2709 [hep-ph]} \BibitemShut
  {NoStop}%
%%CITATION = ARXIV:0709.2709;%%
\bibitem [{\citenamefont {Korchemsky}\ and\ \citenamefont
  {Radyushkin}(1987)}]{Korchemsky:1987wg}%
  \BibitemOpen
  \bibfield  {author} {\bibinfo {author} {\bibfnamefont {G.~P.}\ \bibnamefont
  {Korchemsky}}\ and\ \bibinfo {author} {\bibfnamefont {A.~V.}\ \bibnamefont
  {Radyushkin}},\ }\href@noop {} {\bibfield  {journal} {\bibinfo  {journal}
  {Nucl. Phys. B}\ }\textbf {\bibinfo {volume} {283}},\ \bibinfo {pages} {342}
  (\bibinfo {year} {1987})}\BibitemShut {NoStop}%
%%CITATION = NUPHA,B283,342;%%
\bibitem [{\citenamefont {Moult}\ \emph {et~al.}(2016)\citenamefont {Moult},
  \citenamefont {Stewart}, \citenamefont {Tackmann},\ and\ \citenamefont
  {Waalewijn}}]{Moult:2015aoa}%
  \BibitemOpen
  \bibfield  {author} {\bibinfo {author} {\bibfnamefont {I.}~\bibnamefont
  {Moult}}, \bibinfo {author} {\bibfnamefont {I.~W.}\ \bibnamefont {Stewart}},
  \bibinfo {author} {\bibfnamefont {F.~J.}\ \bibnamefont {Tackmann}}, \ and\
  \bibinfo {author} {\bibfnamefont {W.~J.}\ \bibnamefont {Waalewijn}},\ }\href
  {\doibase 10.1103/PhysRevD.93.094003} {\bibfield  {journal} {\bibinfo
  {journal} {Phys. Rev.}\ }\textbf {\bibinfo {volume} {D93}},\ \bibinfo {pages}
  {094003} (\bibinfo {year} {2016})},\ \Eprint
  {http://arxiv.org/abs/1508.02397} {arXiv:1508.02397 [hep-ph]} \BibitemShut
  {NoStop}%
%%CITATION = ARXIV:1508.02397;%%
\bibitem [{\citenamefont {Balzereit}\ \emph {et~al.}(1998)\citenamefont
  {Balzereit}, \citenamefont {Mannel},\ and\ \citenamefont
  {Kilian}}]{Balzereit:1998yf}%
  \BibitemOpen
  \bibfield  {author} {\bibinfo {author} {\bibfnamefont {C.}~\bibnamefont
  {Balzereit}}, \bibinfo {author} {\bibfnamefont {T.}~\bibnamefont {Mannel}}, \
  and\ \bibinfo {author} {\bibfnamefont {W.}~\bibnamefont {Kilian}},\ }\href
  {\doibase 10.1103/PhysRevD.58.114029} {\bibfield  {journal} {\bibinfo
  {journal} {Phys. Rev.}\ }\textbf {\bibinfo {volume} {D58}},\ \bibinfo {pages}
  {114029} (\bibinfo {year} {1998})},\ \Eprint
  {http://arxiv.org/abs/hep-ph/9805297} {arXiv:hep-ph/9805297} \BibitemShut
  {NoStop}%
%%CITATION = HEP-PH/9805297;%%
\bibitem [{\citenamefont {Neubert}(2005)}]{Neubert:2004dd}%
  \BibitemOpen
  \bibfield  {author} {\bibinfo {author} {\bibfnamefont {M.}~\bibnamefont
  {Neubert}},\ }\href {\doibase 10.1140/epjc/s2005-02141-1} {\bibfield
  {journal} {\bibinfo  {journal} {Eur. Phys. J.}\ }\textbf {\bibinfo {volume}
  {C40}},\ \bibinfo {pages} {165} (\bibinfo {year} {2005})},\ \Eprint
  {http://arxiv.org/abs/hep-ph/0408179} {arXiv:hep-ph/0408179} \BibitemShut
  {NoStop}%
%%CITATION = HEP-PH/0408179;%%
\bibitem [{\citenamefont {Fleming}\ \emph
  {et~al.}(2008{\natexlab{b}})\citenamefont {Fleming}, \citenamefont {Hoang},
  \citenamefont {Mantry},\ and\ \citenamefont {Stewart}}]{Fleming:2007xt}%
  \BibitemOpen
  \bibfield  {author} {\bibinfo {author} {\bibfnamefont {S.}~\bibnamefont
  {Fleming}}, \bibinfo {author} {\bibfnamefont {A.~H.}\ \bibnamefont {Hoang}},
  \bibinfo {author} {\bibfnamefont {S.}~\bibnamefont {Mantry}}, \ and\ \bibinfo
  {author} {\bibfnamefont {I.~W.}\ \bibnamefont {Stewart}},\ }\href {\doibase
  10.1103/PhysRevD.77.114003} {\bibfield  {journal} {\bibinfo  {journal} {Phys.
  Rev.}\ }\textbf {\bibinfo {volume} {D77}},\ \bibinfo {pages} {114003}
  (\bibinfo {year} {2008}{\natexlab{b}})},\ \Eprint
  {http://arxiv.org/abs/0711.2079} {arXiv:0711.2079 [hep-ph]} \BibitemShut
  {NoStop}%
%%CITATION = ARXIV:0711.2079;%%
\bibitem [{\citenamefont {Kramer}\ and\ \citenamefont
  {Lampe}(1987)}]{Kramer:1986sg}%
  \BibitemOpen
  \bibfield  {author} {\bibinfo {author} {\bibfnamefont {G.}~\bibnamefont
  {Kramer}}\ and\ \bibinfo {author} {\bibfnamefont {B.}~\bibnamefont {Lampe}},\
  }\href {\doibase 10.1007/BF01679868} {\bibfield  {journal} {\bibinfo
  {journal} {Z. Phys.}\ }\textbf {\bibinfo {volume} {C34}},\ \bibinfo {pages}
  {497} (\bibinfo {year} {1987})},\ \bibinfo {note} {[Erratum: Z.
  Phys.C42,504(1989)]}\BibitemShut {NoStop}%
%%CITATION = ZEPYA,C34,497;%%
\bibitem [{\citenamefont {Matsuura}\ and\ \citenamefont {van
  Neerven}(1988)}]{Matsuura:1987wt}%
  \BibitemOpen
  \bibfield  {author} {\bibinfo {author} {\bibfnamefont {T.}~\bibnamefont
  {Matsuura}}\ and\ \bibinfo {author} {\bibfnamefont {W.~L.}\ \bibnamefont {van
  Neerven}},\ }\href {\doibase 10.1007/BF01624369} {\bibfield  {journal}
  {\bibinfo  {journal} {Z. Phys.}\ }\textbf {\bibinfo {volume} {C38}},\
  \bibinfo {pages} {623} (\bibinfo {year} {1988})}\BibitemShut {NoStop}%
%%CITATION = ZEPYA,C38,623;%%
\bibitem [{\citenamefont {Matsuura}\ \emph {et~al.}(1989)\citenamefont
  {Matsuura}, \citenamefont {van~der Marck},\ and\ \citenamefont {van
  Neerven}}]{Matsuura:1988sm}%
  \BibitemOpen
  \bibfield  {author} {\bibinfo {author} {\bibfnamefont {T.}~\bibnamefont
  {Matsuura}}, \bibinfo {author} {\bibfnamefont {S.~C.}\ \bibnamefont {van~der
  Marck}}, \ and\ \bibinfo {author} {\bibfnamefont {W.~L.}\ \bibnamefont {van
  Neerven}},\ }\href {\doibase 10.1016/0550-3213(89)90620-2} {\bibfield
  {journal} {\bibinfo  {journal} {Nucl. Phys.}\ }\textbf {\bibinfo {volume}
  {B319}},\ \bibinfo {pages} {570} (\bibinfo {year} {1989})}\BibitemShut
  {NoStop}%
%%CITATION = NUPHA,B319,570;%%
\bibitem [{\citenamefont {Becher}\ \emph {et~al.}(2007)\citenamefont {Becher},
  \citenamefont {Neubert},\ and\ \citenamefont {Pecjak}}]{Becher:2006mr}%
  \BibitemOpen
  \bibfield  {author} {\bibinfo {author} {\bibfnamefont {T.}~\bibnamefont
  {Becher}}, \bibinfo {author} {\bibfnamefont {M.}~\bibnamefont {Neubert}}, \
  and\ \bibinfo {author} {\bibfnamefont {B.~D.}\ \bibnamefont {Pecjak}},\
  }\href {\doibase 10.1088/1126-6708/2007/01/076} {\bibfield  {journal}
  {\bibinfo  {journal} {JHEP}\ }\textbf {\bibinfo {volume} {01}},\ \bibinfo
  {pages} {076} (\bibinfo {year} {2007})},\ \Eprint
  {http://arxiv.org/abs/hep-ph/0607228} {arXiv:hep-ph/0607228} \BibitemShut
  {NoStop}%
%%CITATION = HEP-PH/0607228;%%
\bibitem [{\citenamefont {Idilbi}\ \emph {et~al.}(2006)\citenamefont {Idilbi},
  \citenamefont {Ji},\ and\ \citenamefont {Yuan}}]{Idilbi:2006dg}%
  \BibitemOpen
  \bibfield  {author} {\bibinfo {author} {\bibfnamefont {A.}~\bibnamefont
  {Idilbi}}, \bibinfo {author} {\bibfnamefont {X.-d.}\ \bibnamefont {Ji}}, \
  and\ \bibinfo {author} {\bibfnamefont {F.}~\bibnamefont {Yuan}},\ }\href
  {\doibase 10.1016/j.nuclphysb.2006.07.002} {\bibfield  {journal} {\bibinfo
  {journal} {Nucl. Phys.}\ }\textbf {\bibinfo {volume} {B753}},\ \bibinfo
  {pages} {42} (\bibinfo {year} {2006})},\ \Eprint
  {http://arxiv.org/abs/hep-ph/0605068} {arXiv:hep-ph/0605068} \BibitemShut
  {NoStop}%
%%CITATION = HEP-PH/0605068;%%
\bibitem [{\citenamefont {Harlander}\ and\ \citenamefont
  {Ozeren}(2009)}]{Harlander:2009bw}%
  \BibitemOpen
  \bibfield  {author} {\bibinfo {author} {\bibfnamefont {R.~V.}\ \bibnamefont
  {Harlander}}\ and\ \bibinfo {author} {\bibfnamefont {K.~J.}\ \bibnamefont
  {Ozeren}},\ }\href {\doibase 10.1016/j.physletb.2009.08.012} {\bibfield
  {journal} {\bibinfo  {journal} {Phys. Lett.}\ }\textbf {\bibinfo {volume}
  {B679}},\ \bibinfo {pages} {467} (\bibinfo {year} {2009})},\ \Eprint
  {http://arxiv.org/abs/0907.2997} {arXiv:0907.2997 [hep-ph]} \BibitemShut
  {NoStop}%
%%CITATION = ARXIV:0907.2997;%%
\bibitem [{\citenamefont {Pak}\ \emph {et~al.}(2009)\citenamefont {Pak},
  \citenamefont {Rogal},\ and\ \citenamefont {Steinhauser}}]{Pak:2009bx}%
  \BibitemOpen
  \bibfield  {author} {\bibinfo {author} {\bibfnamefont {A.}~\bibnamefont
  {Pak}}, \bibinfo {author} {\bibfnamefont {M.}~\bibnamefont {Rogal}}, \ and\
  \bibinfo {author} {\bibfnamefont {M.}~\bibnamefont {Steinhauser}},\ }\href
  {\doibase 10.1016/j.physletb.2009.08.016} {\bibfield  {journal} {\bibinfo
  {journal} {Phys. Lett.}\ }\textbf {\bibinfo {volume} {B679}},\ \bibinfo
  {pages} {473} (\bibinfo {year} {2009})},\ \Eprint
  {http://arxiv.org/abs/0907.2998} {arXiv:0907.2998 [hep-ph]} \BibitemShut
  {NoStop}%
%%CITATION = ARXIV:0907.2998;%%
\bibitem [{\citenamefont {Berger}\ \emph {et~al.}(2011)\citenamefont {Berger},
  \citenamefont {Marcantonini}, \citenamefont {Stewart}, \citenamefont
  {Tackmann},\ and\ \citenamefont {Waalewijn}}]{Berger:2010xi}%
  \BibitemOpen
  \bibfield  {author} {\bibinfo {author} {\bibfnamefont {C.~F.}\ \bibnamefont
  {Berger}}, \bibinfo {author} {\bibfnamefont {C.}~\bibnamefont
  {Marcantonini}}, \bibinfo {author} {\bibfnamefont {I.~W.}\ \bibnamefont
  {Stewart}}, \bibinfo {author} {\bibfnamefont {F.~J.}\ \bibnamefont
  {Tackmann}}, \ and\ \bibinfo {author} {\bibfnamefont {W.~J.}\ \bibnamefont
  {Waalewijn}},\ }\href {\doibase 10.1007/JHEP04(2011)092} {\bibfield
  {journal} {\bibinfo  {journal} {JHEP}\ }\textbf {\bibinfo {volume} {04}},\
  \bibinfo {pages} {092} (\bibinfo {year} {2011})},\ \Eprint
  {http://arxiv.org/abs/1012.4480} {arXiv:1012.4480 [hep-ph]} \BibitemShut
  {NoStop}%
%%CITATION = ARXIV:1012.4480;%%
\bibitem [{\citenamefont {Becher}\ and\ \citenamefont
  {Neubert}(2006)}]{Becher:2006qw}%
  \BibitemOpen
  \bibfield  {author} {\bibinfo {author} {\bibfnamefont {T.}~\bibnamefont
  {Becher}}\ and\ \bibinfo {author} {\bibfnamefont {M.}~\bibnamefont
  {Neubert}},\ }\href {\doibase 10.1016/j.physletb.2006.04.046} {\bibfield
  {journal} {\bibinfo  {journal} {Phys. Lett.}\ }\textbf {\bibinfo {volume}
  {B637}},\ \bibinfo {pages} {251} (\bibinfo {year} {2006})},\ \Eprint
  {http://arxiv.org/abs/hep-ph/0603140} {arXiv:hep-ph/0603140} \BibitemShut
  {NoStop}%
%%CITATION = HEP-PH/0603140;%%
\bibitem [{\citenamefont {Becher}\ and\ \citenamefont
  {Bell}(2011)}]{Becher:2010pd}%
  \BibitemOpen
  \bibfield  {author} {\bibinfo {author} {\bibfnamefont {T.}~\bibnamefont
  {Becher}}\ and\ \bibinfo {author} {\bibfnamefont {G.}~\bibnamefont {Bell}},\
  }\href {\doibase 10.1016/j.physletb.2010.11.036} {\bibfield  {journal}
  {\bibinfo  {journal} {Phys. Lett.}\ }\textbf {\bibinfo {volume} {B695}},\
  \bibinfo {pages} {252} (\bibinfo {year} {2011})},\ \Eprint
  {http://arxiv.org/abs/1008.1936} {arXiv:1008.1936 [hep-ph]} \BibitemShut
  {NoStop}%
%%CITATION = ARXIV:1008.1936;%%
\bibitem [{\citenamefont {Kelley}\ \emph {et~al.}(2011)\citenamefont {Kelley},
  \citenamefont {Schwartz}, \citenamefont {Schabinger},\ and\ \citenamefont
  {Zhu}}]{Kelley:2011ng}%
  \BibitemOpen
  \bibfield  {author} {\bibinfo {author} {\bibfnamefont {R.}~\bibnamefont
  {Kelley}}, \bibinfo {author} {\bibfnamefont {M.~D.}\ \bibnamefont
  {Schwartz}}, \bibinfo {author} {\bibfnamefont {R.~M.}\ \bibnamefont
  {Schabinger}}, \ and\ \bibinfo {author} {\bibfnamefont {H.~X.}\ \bibnamefont
  {Zhu}},\ }\href {\doibase 10.1103/PhysRevD.84.045022} {\bibfield  {journal}
  {\bibinfo  {journal} {Phys. Rev.}\ }\textbf {\bibinfo {volume} {D84}},\
  \bibinfo {pages} {045022} (\bibinfo {year} {2011})},\ \Eprint
  {http://arxiv.org/abs/1105.3676} {arXiv:1105.3676 [hep-ph]} \BibitemShut
  {NoStop}%
%%CITATION = ARXIV:1105.3676;%%
\bibitem [{\citenamefont {Hornig}\ \emph {et~al.}(2011)\citenamefont {Hornig},
  \citenamefont {Lee}, \citenamefont {Stewart}, \citenamefont {Walsh},\ and\
  \citenamefont {Zuberi}}]{Hornig:2011iu}%
  \BibitemOpen
  \bibfield  {author} {\bibinfo {author} {\bibfnamefont {A.}~\bibnamefont
  {Hornig}}, \bibinfo {author} {\bibfnamefont {C.}~\bibnamefont {Lee}},
  \bibinfo {author} {\bibfnamefont {I.~W.}\ \bibnamefont {Stewart}}, \bibinfo
  {author} {\bibfnamefont {J.~R.}\ \bibnamefont {Walsh}}, \ and\ \bibinfo
  {author} {\bibfnamefont {S.}~\bibnamefont {Zuberi}},\ }\href {\doibase
  10.1007/JHEP08(2011)054} {\bibfield  {journal} {\bibinfo  {journal} {JHEP}\
  }\textbf {\bibinfo {volume} {08}},\ \bibinfo {pages} {054} (\bibinfo {year}
  {2011})},\ \Eprint {http://arxiv.org/abs/1105.4628} {arXiv:1105.4628
  [hep-ph]} \BibitemShut {NoStop}%
%%CITATION = ARXIV:1105.4628;%%
\bibitem [{\citenamefont {Tarasov}\ \emph {et~al.}(1980)\citenamefont
  {Tarasov}, \citenamefont {Vladimirov},\ and\ \citenamefont
  {Zharkov}}]{Tarasov:1980au}%
  \BibitemOpen
  \bibfield  {author} {\bibinfo {author} {\bibfnamefont {O.~V.}\ \bibnamefont
  {Tarasov}}, \bibinfo {author} {\bibfnamefont {A.~A.}\ \bibnamefont
  {Vladimirov}}, \ and\ \bibinfo {author} {\bibfnamefont {A.~{\relax Yu}.}\
  \bibnamefont {Zharkov}},\ }\href {\doibase 10.1016/0370-2693(80)90358-5}
  {\bibfield  {journal} {\bibinfo  {journal} {Phys. Lett.}\ }\textbf {\bibinfo
  {volume} {B93}},\ \bibinfo {pages} {429} (\bibinfo {year}
  {1980})}\BibitemShut {NoStop}%
%%CITATION = PHLTA,B93,429;%%
\bibitem [{\citenamefont {Larin}\ and\ \citenamefont
  {Vermaseren}(1993)}]{Larin:1993tp}%
  \BibitemOpen
  \bibfield  {author} {\bibinfo {author} {\bibfnamefont {S.~A.}\ \bibnamefont
  {Larin}}\ and\ \bibinfo {author} {\bibfnamefont {J.~A.~M.}\ \bibnamefont
  {Vermaseren}},\ }\href {\doibase 10.1016/0370-2693(93)91441-O} {\bibfield
  {journal} {\bibinfo  {journal} {Phys. Lett.}\ }\textbf {\bibinfo {volume}
  {B303}},\ \bibinfo {pages} {334} (\bibinfo {year} {1993})},\ \Eprint
  {http://arxiv.org/abs/hep-ph/9302208} {arXiv:hep-ph/9302208} \BibitemShut
  {NoStop}%
%%CITATION = HEP-PH/9302208;%%
\bibitem [{\citenamefont {Moch}\ \emph {et~al.}(2004)\citenamefont {Moch},
  \citenamefont {Vermaseren},\ and\ \citenamefont {Vogt}}]{Moch:2004pa}%
  \BibitemOpen
  \bibfield  {author} {\bibinfo {author} {\bibfnamefont {S.}~\bibnamefont
  {Moch}}, \bibinfo {author} {\bibfnamefont {J.~A.~M.}\ \bibnamefont
  {Vermaseren}}, \ and\ \bibinfo {author} {\bibfnamefont {A.}~\bibnamefont
  {Vogt}},\ }\href {\doibase 10.1016/j.nuclphysb.2004.03.030} {\bibfield
  {journal} {\bibinfo  {journal} {Nucl. Phys.}\ }\textbf {\bibinfo {volume}
  {B688}},\ \bibinfo {pages} {101} (\bibinfo {year} {2004})},\ \Eprint
  {http://arxiv.org/abs/hep-ph/0403192} {arXiv:hep-ph/0403192} \BibitemShut
  {NoStop}%
%%CITATION = HEP-PH/0403192;%%
\bibitem [{\citenamefont {Becher}\ and\ \citenamefont
  {Schwartz}(2010)}]{Becher:2009th}%
  \BibitemOpen
  \bibfield  {author} {\bibinfo {author} {\bibfnamefont {T.}~\bibnamefont
  {Becher}}\ and\ \bibinfo {author} {\bibfnamefont {M.~D.}\ \bibnamefont
  {Schwartz}},\ }\href {\doibase 10.1007/JHEP02(2010)040} {\bibfield  {journal}
  {\bibinfo  {journal} {JHEP}\ }\textbf {\bibinfo {volume} {02}},\ \bibinfo
  {pages} {040} (\bibinfo {year} {2010})},\ \Eprint
  {http://arxiv.org/abs/0911.0681} {arXiv:0911.0681 [hep-ph]} \BibitemShut
  {NoStop}%
%%CITATION = ARXIV:0911.0681;%%
\bibitem [{\citenamefont {Ellis}\ \emph {et~al.}(1981)\citenamefont {Ellis},
  \citenamefont {Ross},\ and\ \citenamefont {Terrano}}]{Ellis:1980wv}%
  \BibitemOpen
  \bibfield  {author} {\bibinfo {author} {\bibfnamefont {R.~K.}\ \bibnamefont
  {Ellis}}, \bibinfo {author} {\bibfnamefont {D.~A.}\ \bibnamefont {Ross}}, \
  and\ \bibinfo {author} {\bibfnamefont {A.~E.}\ \bibnamefont {Terrano}},\
  }\href {\doibase 10.1016/0550-3213(81)90165-6} {\bibfield  {journal}
  {\bibinfo  {journal} {Nucl. Phys.}\ }\textbf {\bibinfo {volume} {B178}},\
  \bibinfo {pages} {421} (\bibinfo {year} {1981})}\BibitemShut {NoStop}%
%%CITATION = NUPHA,B178,421;%%
\bibitem [{\citenamefont {Schmidt}(1997)}]{Schmidt:1997wr}%
  \BibitemOpen
  \bibfield  {author} {\bibinfo {author} {\bibfnamefont {C.~R.}\ \bibnamefont
  {Schmidt}},\ }\href {\doibase 10.1016/S0370-2693(97)01102-7} {\bibfield
  {journal} {\bibinfo  {journal} {Phys. Lett.}\ }\textbf {\bibinfo {volume}
  {B413}},\ \bibinfo {pages} {391} (\bibinfo {year} {1997})},\ \Eprint
  {http://arxiv.org/abs/hep-ph/9707448} {arXiv:hep-ph/9707448} \BibitemShut
  {NoStop}%
%%CITATION = HEP-PH/9707448;%%
\bibitem [{\citenamefont {Bernlochner}\ \emph {et~al.}(2013)\citenamefont
  {Bernlochner}, \citenamefont {Lacker}, \citenamefont {Ligeti}, \citenamefont
  {Stewart}, \citenamefont {Tackmann},\ and\ \citenamefont
  {Tackmann}}]{Bernlochner:2013gla}%
  \BibitemOpen
  \bibfield  {author} {\bibinfo {author} {\bibfnamefont {F.~U.}\ \bibnamefont
  {Bernlochner}}, \bibinfo {author} {\bibfnamefont {H.}~\bibnamefont {Lacker}},
  \bibinfo {author} {\bibfnamefont {Z.}~\bibnamefont {Ligeti}}, \bibinfo
  {author} {\bibfnamefont {I.~W.}\ \bibnamefont {Stewart}}, \bibinfo {author}
  {\bibfnamefont {F.~J.}\ \bibnamefont {Tackmann}}, \ and\ \bibinfo {author}
  {\bibfnamefont {K.}~\bibnamefont {Tackmann}} (\bibinfo {collaboration}
  {SIMBA}),\ }\bibfield  {booktitle} {\emph {\bibinfo {booktitle} {{7th
  International Workshop on the CKM Unitarity Triangle (CKM 2012) Cincinnati,
  Ohio, USA, September 28-October 2, 2012}}},\ }\href@noop {} {\  (\bibinfo
  {year} {2013})},\ \bibinfo {note} {[PoSICHEP2012,370(2013)]},\ \Eprint
  {http://arxiv.org/abs/1303.0958} {arXiv:1303.0958 [hep-ph]} \BibitemShut
  {NoStop}%
%%CITATION = ARXIV:1303.0958;%%
\bibitem [{\citenamefont {Stewart}\ \emph {et~al.}(2015)\citenamefont
  {Stewart}, \citenamefont {Tackmann},\ and\ \citenamefont
  {Waalewijn}}]{Stewart:2014nna}%
  \BibitemOpen
  \bibfield  {author} {\bibinfo {author} {\bibfnamefont {I.~W.}\ \bibnamefont
  {Stewart}}, \bibinfo {author} {\bibfnamefont {F.~J.}\ \bibnamefont
  {Tackmann}}, \ and\ \bibinfo {author} {\bibfnamefont {W.~J.}\ \bibnamefont
  {Waalewijn}},\ }\href {\doibase 10.1103/PhysRevLett.114.092001} {\bibfield
  {journal} {\bibinfo  {journal} {Phys. Rev. Lett.}\ }\textbf {\bibinfo
  {volume} {114}},\ \bibinfo {pages} {092001} (\bibinfo {year} {2015})},\
  \Eprint {http://arxiv.org/abs/1405.6722} {arXiv:1405.6722 [hep-ph]}
  \BibitemShut {NoStop}%
%%CITATION = ARXIV:1405.6722;%%
\bibitem [{\citenamefont {Stewart}\ \emph {et~al.}(2014)\citenamefont
  {Stewart}, \citenamefont {Tackmann}, \citenamefont {Walsh},\ and\
  \citenamefont {Zuberi}}]{Stewart:2013faa}%
  \BibitemOpen
  \bibfield  {author} {\bibinfo {author} {\bibfnamefont {I.~W.}\ \bibnamefont
  {Stewart}}, \bibinfo {author} {\bibfnamefont {F.~J.}\ \bibnamefont
  {Tackmann}}, \bibinfo {author} {\bibfnamefont {J.~R.}\ \bibnamefont {Walsh}},
  \ and\ \bibinfo {author} {\bibfnamefont {S.}~\bibnamefont {Zuberi}},\ }\href
  {\doibase 10.1103/PhysRevD.89.054001} {\bibfield  {journal} {\bibinfo
  {journal} {Phys. Rev.}\ }\textbf {\bibinfo {volume} {D89}},\ \bibinfo {pages}
  {054001} (\bibinfo {year} {2014})},\ \Eprint {http://arxiv.org/abs/1307.1808}
  {arXiv:1307.1808 [hep-ph]} \BibitemShut {NoStop}%
%%CITATION = ARXIV:1307.1808;%%
\bibitem [{\citenamefont {Gangal}\ \emph {et~al.}(2015)\citenamefont {Gangal},
  \citenamefont {Stahlhofen},\ and\ \citenamefont {Tackmann}}]{Gangal:2014qda}%
  \BibitemOpen
  \bibfield  {author} {\bibinfo {author} {\bibfnamefont {S.}~\bibnamefont
  {Gangal}}, \bibinfo {author} {\bibfnamefont {M.}~\bibnamefont {Stahlhofen}},
  \ and\ \bibinfo {author} {\bibfnamefont {F.~J.}\ \bibnamefont {Tackmann}},\
  }\href {\doibase 10.1103/PhysRevD.91.054023} {\bibfield  {journal} {\bibinfo
  {journal} {Phys. Rev.}\ }\textbf {\bibinfo {volume} {D91}},\ \bibinfo {pages}
  {054023} (\bibinfo {year} {2015})},\ \Eprint {http://arxiv.org/abs/1412.4792}
  {arXiv:1412.4792 [hep-ph]} \BibitemShut {NoStop}%
%%CITATION = ARXIV:1412.4792;%%
\end{thebibliography}%

\end{document}